\documentstyle[11pt,aaspp4]{article}

\begin{document}
\title{CANADA-FRANCE REDSHIFT SURVEY XIV:\\
SPECTRAL PROPERTIES OF FIELD GALAXIES UP TO z = 1}

\author{F. Hammer$^{a,1}$, H. Flores$^{a,b}$, S. J. Lilly$^{c,1}$, 
David Crampton$^{d,1}$, O. Le F\`evre$^{a,1}$, C. Rola$^e$,
G. Mallen-Ornelas$^c$, D. Schade$^c,d$, L. Tresse$^e$}

\affil{a) Observatoire de Paris, Section de Meudon, DAEC, 92195 Meudon Principal Cedex, France.\\
b) Universitad Catolica de Chile, Santiago, Chile.\\ 
c) Department of Astronomy, University of Toronto, Toronto, Canada.\\
d) Dominion Astrophysical Observatory, National Research Council of Canada, Victoria, Canada\\
e) Institute of Astronomy, Cambridge, UK.}
\authoremail{hammer@obspm.fr}

\altaffiltext{1} {Visiting Astronomer, Canada-France-Hawaii Telescope,
which is operated by the National Research Council of Canada, the
Centre National de la Recherche Scientifique, and the University of
Hawaii}

\begin{abstract}

The spectral properties of more than 400 CFRS galaxies and their changes
over the redshift interval $0 \le z \le 1.3$ are investigated.
Emission line intensities and equivalent widths for accessible lines
have been measured, as well as continuum color indices based on
200\AA\ wide spectral regions.  Within the CFRS sample, the  
comoving fraction
of galaxies with significant emission lines ($W_{0}(OII) >$ 15\AA)
increases from $\sim13\%$ locally to over 50\% at z $>$ 0.5. The
fraction of luminous ( $M_{B} <$ -20) quiescent galaxies (those without [OII] 3727
emission) decreases with redshift from 53\% at z = 0.3 to 23\% at
z$>$0.5, the latter fraction being similar to that of early type
galaxies at that redshift.  There is considerable evidence in the data
presented here that star formation increases from z = 0 to z $>$ 0.5
in disk galaxies.  However, the absence of extremely blue colors and
the presence of significant Balmer absorption suggests that the star
formation is primarily taking place over long periods of time, rather
than in short-duration, high-amplitude ``bursts''.

There are several indications that the average metallicity and dust
opacity were lower in emission-line galaxies at high redshift than
those typically seen in luminous galaxies locally.  Beyond z = 0.7,
almost all the emission-line galaxies, including the most luminous
(at 1$\mu$ at rest) ones, have colors approaching those of present-day
 irregular galaxies,
and a third of them have indications (primarily from the strength of
the 4000\AA\ break) of metallicities significantly less than solar ($Z
< 0.2 Z_{\odot}$).  It is argued that changes in metallicity and dust
extinction could be contributing to the observed evolution of the line
and continuum luminosity densities, the luminosity function and/or the
surface brightnesses and morphologies of galaxies in the CFRS.

If the Kennicutt (1992) relation is used to convert the large increase
in the comoving luminosity density of [OII] 3727 back to z $\sim$ 1
into a star formation rate, it implies a present stellar mass density
in excess of that observed locally. This result suggests that the
Kennicutt relation is inappropriate for the CFRS objects, perhaps due
to the changes in their metallicities and dust opacities, 
 and/or in their IMFs. Using the
Gallagher et al. (1989) relation for objects having colors of
irregulars reduces the production of long-lived stars since z = 1 to
75\% of the present-day value. More complex mechanisms are probably
responsible for changes seen in the emission line ratios of HII
regions in CFRS galaxies, which show higher ionization parameters than
local 
 HII galaxy ones.  This change could be due to a higher ionizing 
efficiency of the photons from hot stars in galaxies at high redshift.

\end{abstract}

\keywords{Galaxies: abundances, evolution, stellar content;  Cosmology: observations}

\section{INTRODUCTION}

The Canada-France Redshift Survey (CFRS) has produced a unique sample
of 730 field galaxies with $I_{AB}<$ 22.5, 591 of which have measured
redshifts in the range $0 < z < 1.4$ (Lilly et al. 1995a, hereafter
CFRS I; Le F\`evre et al. 1995, hereafter CFRS II; Lilly et al. 1995b,
hereafter CFRS III; Hammer et al. 1995a, hereafter CFRS IV). The
spectroscopic incompleteness is low (15$\%$ of success rate in
identifying galaxy and star spectra , Crampton et al. 1995, hereafter
CFRS V), and deep imaging in $I$ was carried out for all the sample in
order to avoid selection biases against low surface brightness galaxies
(see Figure 7 in CFRS I). Selecting objects in the $I$ band
substantially reduces the problems of working at high redshift,
including minimizing {\it k} corrections up to z = 0.9, where the $I$
light corresponds to the $B$ band at rest.  Deep $B, V$, $I$ and $K$
photometry is also available for most of the galaxies ($V$ and $I_{AB}$
in CFRSII,III and IV; $K_{AB}$ in Table 1 of this paper).  This sample
was primarily aimed at a determination of the galaxy luminosity
function up to $z \sim 1$ (see Lilly et al. 1995c; hereafter CFRS VI).
The main result is that there is a change in the luminosity function of
blue galaxies, by one magnitude at z = 0.62 relative to z = 0.375
assuming pure luminosity evolution, while the luminosity function of
red field galaxies shows no evidence for change back to $z \sim 1$ (see
CFRS VI). A further brightening of the luminosity function of blue
galaxies by about 1 mag is found between $0.62 < z < 0.85$. Several
CFRS galaxies have now been observed with $HST$ allowing deconvolution
of the bulge and disk components. Compared to z = 0.3, disks appear to
be brighter in surface brightness by $\sim 1$ magnitude at z = 0.8
(Schade et al. 1995, hereafter CFRS IX, see also Schade et al. 1996b,
hereafter CFRS XI), an effect which likely makes a substantial
contribution to the changes observed in the luminosity function. The
comoving luminosity densities from z = 0.2 to z = 1 in three wave bands
(2800\AA, 4400\AA\ and 10000\AA\ at rest) have been computed from
$BVIK$ photometry (Lilly et al. 1996, hereafter CFRS XIII).  They
increase strongly with redshift at all wavelengths, especially at
2800\AA. Evolution of the spatial correlation function is also observed
(Le F\`evre et al. 1996, hereafter CFRS VIII).

Given this evidence for evolution of the galaxy population, analysis of
the spectral properties of the CFRS galaxies may produce
insights into the physical processes responsible.  It is likely that
galaxy evolution is a complex process involving the evolution of
individual stars, evolution of physical properties such as dust
extinction and metallicity, and other mechanisms operating within the
galaxies and between galaxies and their environments. A
spectrophotometric study of a sample like the CFRS aims to improve the
understanding of the physical processes involved in the observed
evolution, partially through a systematic comparison with well-known
local objects.  Extensive studies of local galaxies have been carried
out by Kennicutt (1992, hereafter K92) and Kennicutt et al. (1994),
although they were not based on complete samples with simple selection
criteria. These studies have provided consistent information about the
present and past rates of star formation, as well as several
relationships between emission lines, continuum and extinction
effects.  There are magnitude-selected samples of local galaxies which
can be compared to the CFRS, including the Las Campanas survey
(Schectman et al. 1992), the DARS sample (Peterson et al. 1986) and the
ESP sample (Vettolani et al. 1996), although some of these have
significant surface-brightness selection effects.  Moreover, to our
knowledge, no systematic studies of the spectrophotometric properties
of galaxies have been carried out for these samples.

Age and metallicity effects can be investigated through comparison with
young, intermediate and old stellar cluster populations.  In their
pioneering work, Bica and Alloin (1986) have studied such effects using
star clusters of various ages and metallicities. The comparison of
spectrophotometric data with predictions from population synthesis
models (Bruzual \& Charlot 1993, 1995, hereafter BC93 and BC95;
Guiderdoni \& Rocca 1987) are also valuable.  These provide evolution
of the stellar tracks for various scenarios, while also helping to
investigate possible effects related to extinction or to abundances
(Worthey 1994; Leitherer and Heckman 1995; Charlot 1996a; Charlot et al, 1996).
Photoionization codes (McCall et al.  1986; Stasinska 1980; Rola 1995)
are useful in investigating the underlying physics of the HII regions.

We present here an analysis of the spectral properties of the CFRS
galaxies. As noted above, useful reference papers are CFRS VI
(luminosity function), CFRS IX and XI (galaxy morphologies) and CFRS
XIII (comoving luminosity density).  Restframe magnitudes and colors
computed as in CFRS VI and CFRS XIII are used for intercomparison of
spectral properties. In section 2, we present the observational data
(see also Table 1) and compare spectral and photometric properties. 
In section 3, the main
spectral properties of CFRS galaxies and their evolution with 
redshift are presented, while the material in section 4 focuses on the
emission line properties of z$< 0.7$ galaxies (HII regions and AGNs
from diagnostic diagrams). Section 5 presents the continuum properties
of galaxies up to $z = 1$ and slightly beyond, and investigates their
underlying physical properties (stellar content and age, extinction and
metallicity).  Sections 6, 7 and 8 describe the properties of several
categories of CFRS galaxies (sub-solar metallicity, quiescent and
peculiar galaxies respectively) and a general discussion is presented
in Section 9. Values of $q_{0}$=0.5, and $h_{50} = 1$ (with $H_{0} = 50
h_{50}$km s$^{-1}$Mpc$^{-1}$) have been adopted throughout the paper,
except where noted to the contrary.

\section{BASIC DATA}

The spectroscopic data were taken at CFHT during several runs, but the
same observational strategy was used throughout, and the CFRS spectra
can be regarded as a homogeneous set of data. The spectroscopic data
are described in detail in CFRS II, III, IV and V. The spectral
resolution is 40\AA\ and the spectral range is 4250 -- 8500\AA. The
spectra  were flux calibrated using independent calibrating stars for
each run (generally 10\% accurate). The CFRS sample has been found to
be free of significant observational selection effects against low
surface brightness or compact objects (CFRS I, CFRS IV, CFRS V). Added
to the three-fold reduction and redshift estimation and to the careful
treatment of instrumental defects (CFRS II), this has ensured that CFRS
galaxies are essentially selected by their magnitudes (17.5 $< I_{AB} <
$ 22.5).  Special care has been taken to provide the highest quality
spectrum for each individual object, selecting the best spectrum from
three independent reductions of the data.  The CFRS currently provides
the largest well-defined sample of galaxy spectra extending out to $z =
1$ that is suitable for spectral analysis.

\subsection{Line measurements}

Line measurements have been made using the software MEASURE implemented
at the Meudon Observatory by D. Pelat. Lines with low S/N ($<$ 2-3)
were not taken into  account (see Rola \& Pelat 1994) and the software
provides an elegant way to estimate the error associated with the
placement of the continuum. Indeed, the latter can be determined either
by the user or by the software ( poissonian noise), and the final error
was set to be the maximum of the error given by the software (gaussian
fit), or the difference between the user and software approximations.
Several emission and absorption lines have been measured in the 591
galaxies including [OII]3727, [OIII]5007, [SII] 6724, $H\delta$,
H$\beta$, $H\alpha$, CaII 3933 and 3969, Mgb 5175 and NaD 5892 when
available in the spectral window. Because the spectral resolution was
very low, lines such as [SII]6716 and [SII]6728 and $H\alpha$ and
[NII]6583 could not be resolved.  Emission line intensities (in units
of $10^{-29}$ erg-\AA $cm^{-2}$) and equivalent widths (\AA) and their
corresponding errors are shown in Table 1.  When a line doesn't appear
in the spectral range, we denote it by 9999, while when the line was
not measurable because of instrumental reasons (bad sky subtraction or
zero order, see CFRS II) it is represented by 9998.  In general,
measurements of absorption lines are very noisy due to the combination
of our poor spectral resolution and the background sky noise. We were
unable to provide absorption line measurements for more than half of
the objects, which led us to use medium-band spectrophotometry to
investigate the spectral properties of CFRS galaxies.

\subsection{Continuum indices}

Six continuum indices have been calculated which, added to the
photometric colors, provide a fair representation of the galaxy
restframe continuum properties. Unlike broad-band photometric
colors, each of the continuum indices can be tailored to one or several
physical properties in the galaxies. We have mapped the continua from
the UV to the visible with:\\
-- a UV color index D(3250-3550) = 
$f_{\nu}$(3450-3650)/$f_{\nu}$(3150-3350),
the ratio of the average flux density $f_{\nu}$ in the bands 
3450-3650\AA\ and
3150-3350\AA\ at rest, which is mostly 
sensitive to hot stars.\\
-- a Balmer index, D(3550-3850) = 
$f_{\nu}$(3750-3950)/$f_{\nu}$(3450-3650),
which is very sensitive to the population of A stars.\\
-- a UV-B index, D(3550-4150), which is the ratio of the average flux density $f_{\nu}$
in the bands 4050-4250\AA\ and 3450-3650\AA\ at rest.\\
-- the 4000\AA\ break index D(4000) as defined by Bruzual (1983), which is
the ratio of the average flux density $f_{\nu}$ in the bands 4050-4250\AA\
and 3750-3950\AA\ at rest; this index depends on both metallicity and
stellar ages.\\
-- the ``red" color index D(41-50)$=
$2.5log($f_{\nu}$(4900-5100)/$f_{\nu}$(4050-4250)),
which follows the definition of K92 and which is rather
independent of the hot star population (except for the youngest
galaxies), while it varies with metallicity.\\
-- the Mg$_2$ index which is defined following Faber et al. (1977) and
Buzzoni et al. (1992),
Mg$_2$ = $-$2.5log(f(5156-5197.25\AA)/ (0.39f(4897-4958.25)+0.61f(5303-5366.75)),
where all the flux densities correspond to average values in the
corresponding bands, either centered on the Mg feature or at the blue
and red sides respectively; this index depends on both temperature and
metallicity (Faber et al. 1977; Faber et al. 1985; Davies et al.
1987).

In order to properly calculate the average flux in a given spectral
band, our software  fit the continuum by a straight line, rejecting
spikes due to bad sky subtraction, residual cosmic rays or possible
emission lines using the method of sigma clipping.  Corresponding
errors have been calculated from the standard deviation. The
7600\AA\ atmospheric $O_{2}$ absorption ($\sim$ 40\AA\ wide) region was excluded
prior to the index calculations. Independent estimates were made (by
H.F., F.H. and G.M-O.) for the D(4000) index and are found to be in
good agreement. Values of the indices and corresponding errors are
shown in Table 1. When an index could not be properly computed because
the corresponding restframe continuum bands lie outside the spectral
range, a value of 9999 is given. In these calculations, we have
restricted the spectral range to 4750-8250\AA, in order to avoid
possible effects near the ends of the spectra.  The available redshift
range is 0.45 $<z<$ 1.3 for the UV index D(3250-3550), 0.33 $<z<$ 1.14
for the Balmer index D(3550-3850), 0.23 $<z<$ 1.0 for the D(4000) index,
 0.15 $<z<$ 0.65 for the  D(41-50) index
 and 0 $<$z$<$ 0.58 for the Mg$_2$ index.

\subsection{Comparison of spectrophotometry and photometry}

Several problems can alter the reliability of the spectrophotometric
calibration for such faint objects, including changes in atmospheric
conditions during the night, some poorly-determined slit profiles (due
to badly cut slits) and poor extraction of the spectrum at the noisier
blue or red ends. Our imaging data was sufficiently deep in broad-band V and
I to allow accurate photometry ( generally better than 0.1 magnitude)
for the $I_{AB}< 22$ CFRS galaxies (CFRS I).  Figure 1 compares the
$(V-I)_{AB}$ index derived from galaxy spectra with that derived from
photometry. For the former, the $I_{AB}$ values have been estimated
assuming that beyond 8500\AA, the limit of our spectroscopy, the slope
of the spectra was the same as that below 8500\AA. The dispersion
between the two indices is 0.15 mag, after excluding the most
discrepant objects  ($\sim$ 30$\%$ of the sample). On average, the
spectroscopic ($V-I$) is slightly redder (0.04 mag) than the photometric
($V-I$), otherwise no systematic trends are apparent in Figure 1.
\begin{figure}[tbp] \label{f1}
\plotone{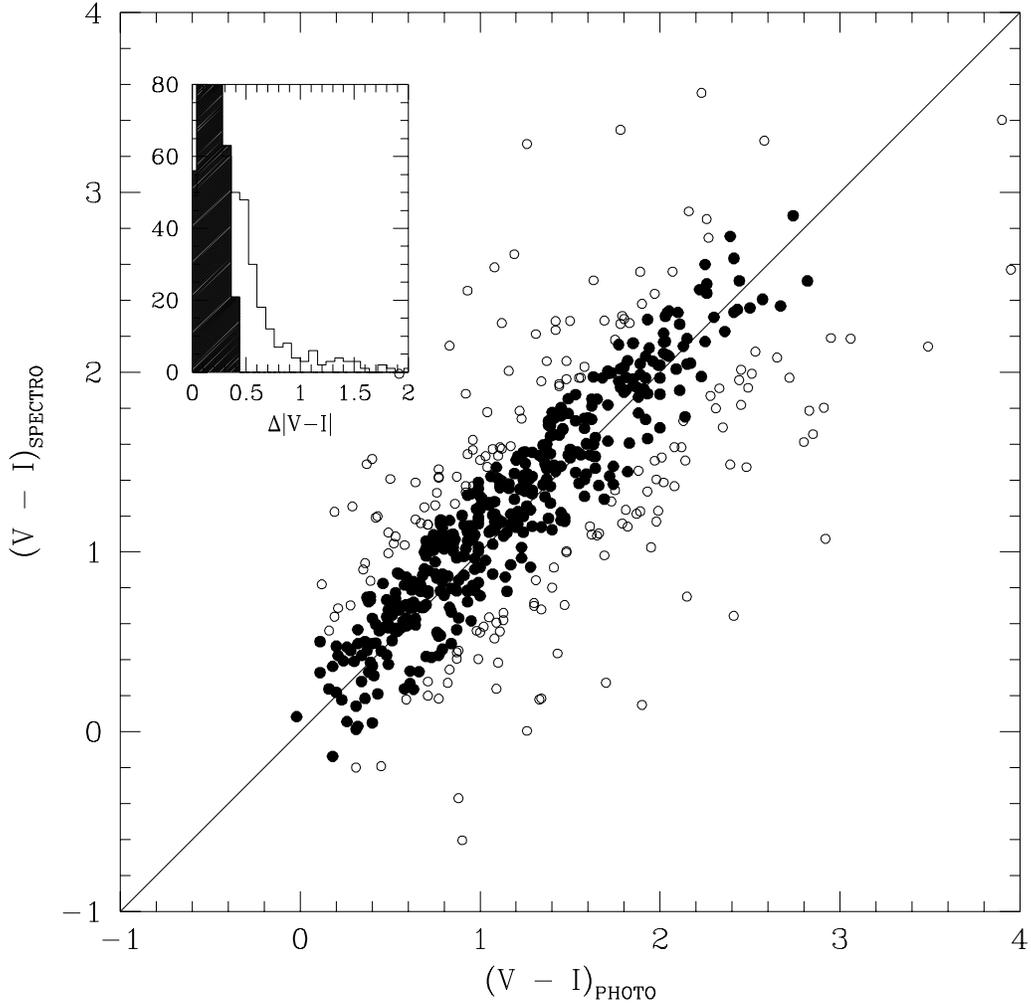}
\caption[]{Plot of the colors derived from the spectra compared to
photometric colors. The residual dispersion of the relation is 0.15
magnitude. Only the 410 galaxies with agreement better than 0.4 mag (full dots,
sample B)
are considered here, as well as in the discussion in the text.  The objects
which are more discrepant than this (open dots) show no systematic peculiarities
compared to the whole sample but, for most of them, the flux
calibration obviously failed over part of the wavelength region. The box in 
the upper left corner shows the distribution of the difference between the
two color estimates.}
\end{figure}

\begin{figure}[tbp] \label{f2}
\plotone{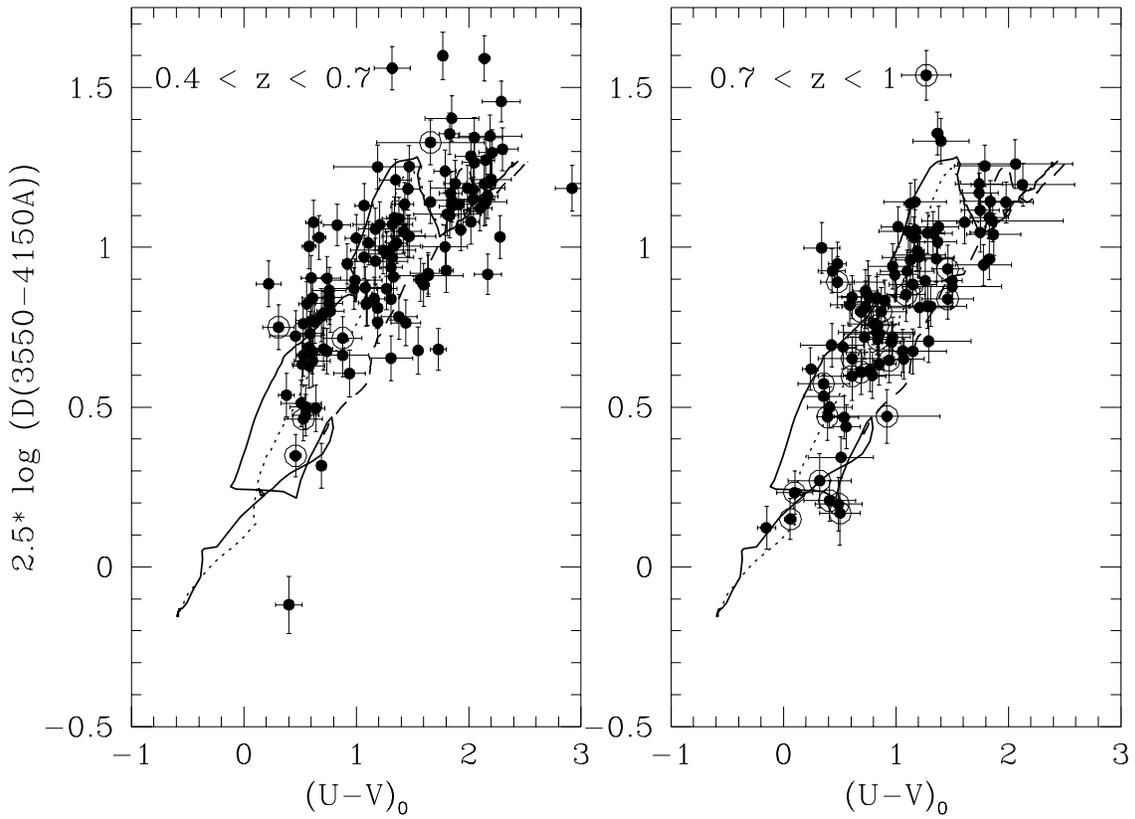}
\caption[]{Plot of the spectral index D(3550-4150) (in magnitude)
compared to the restframe $(U-V)_{AB}$ color derived from the
photometry. Lines show the predictions from BC95 (solid: instantaneous
burst, dotted: 1 Gyr burst, dashed: 10\% strength burst occuring in a 
15 Gyr galaxy).
Almost all the galaxies (sample B) are found near or within an area 
delimited by
solid and dashed lines in both redshift ranges. Note that the
presence of sky emission lines above 7200\AA\ has not seriously
affected the calculation of indices such as D(4000) beyond z = 0.7 or
D(41-50) beyond z = 0.45. Circled dots distinguish D(4000) deficient
objects.}
\end{figure}

\begin{deluxetable}{cccc}
\tablenum{2}
\tablewidth{0pt}
\tablecaption{Sample and sub-samples used in the paper}
\tablehead{ 
\colhead{Sample} & \colhead{object number} & 
\colhead{subsample of} & \colhead{criterion} 
}
\startdata
A & 591 &   & CFRS galaxies, 0 $<$ z $<$ 1.4                 \nl
B & 410 & A & $|(V-I)_{PHOTO} - (V-I)_{SPECTRO}| <$ 0.4 mag  \nl
C & 272 & B & $M_{B}<$ -20 galaxies and [OII] 3727 measurable\nl
D & 212 & C & $W_{0}(OII)>$ 0                                \nl
E & 232 & C & 0.35$<$z$<$ 1 (and D(3550-4150 measurable)     \nl
F & 132 & A & $M_{1\mu}<$ -22 and [OII] 3727 measurable       \nl
G & 83  & A & -22$< M_{1\mu}<$ -21 and [OII] 3727 measurable   \nl
H & 162 & A & [OII] 3727, [OIII] 5007 and D(3550-3850) measurable\nl
I & 102 & H & $W_{0}(OII)>$ 0 and $W_{0}(OIII)>$ 0           \nl
\enddata
\end{deluxetable}

\subsection{Definition of subsamples}

In the following, only the 410 galaxies (of 591 in the CFRS complete
sample) for which the spectroscopic and photometric ($V-I$) colors agree
to within 0.4 mag (see Figure 1) are considered (sample B). Limiting the 
sample in
this way minimizes any instrumental biases related to our
spectroscopy in determining the continuum indices and their reliability
from one spectrum to another, emission line ratios (especially those
lying in different parts of the spectrum), and the fits of individual
spectra by model templates.  A 0.4 mag discrepancy in $(V-I)_{AB}$
would lead to an uncertainty of the D(4000) index of 0.07 mag (or 0.03
in log scale), for example, if the discrepancy propagates linearly
with wavelength. Other sources of errors arise from the index or
line measurements and from the photometric color measurements (CFRS
I). The relationship between the D(3550-4150) spectral index and the
rest frame $(U-V)_{AB}$ color (estimated from the photometry) suggests
that the spectral index accuracies are comparable to those of the
restframe photometric colors, for almost all galaxies (Figure 2). This
is also valid for z = 0.7 to z = 1 galaxies, for which the
4050-4250\AA\ band is redshifted beyond 7100\AA, a spectral
range dominated by sky emission lines. 

Table 2 presents the various sub-samples used in the paper. 
In much of the discussion we include only those galaxies with $M_{B}<
-20$, since less luminous objects rapidly slip beyond the survey limit
for $z > 0.5$ (samples C, D and E). The sample is frequently 
subdivided into two luminosity ranges: a luminous bin ($M_{1\mu} < -22$,
sample F) and a moderately luminous bin ($-22 < M_{1\mu} <-21$, sample G),
 which allows intercomparison of galaxy properties from z = 0.4
to z = 1.1, and from z = 0.3 to z = 0.9 respectively. Luminosities at
1$\mu$ (rest wavelength, monochromatic) have been calculated from the $I$ and 
$K$ photometry (see CFRS XIII). Values for $K_{AB}$ and
for $M_{1\mu}$ are provided in Table 1, latter could be used by the 
reader to define the sub-samples described along the paper.
The luminosity at 1$\mu$ at all redshifts is more closely
related to the amount of old stars in galaxies than the B luminosity
and hence, although in a complicated way, to their masses. Other subdivisions
concerns the emission line objects and their classifications (samples H and
I). 


\subsubsection{Quiescent galaxies}

We use this generic term for all galaxies without detected emission
lines. It must be kept in mind that a significant fraction of quiescent
galaxies may not have been identified at z $>$ 0.7 (see below and
CFRS V) and also
that our spectral range does not include H$\alpha$ at z $>$ 0.3 and
H$\beta$ and [OIII] 4959,5007  beyond z~=~0.7.

\begin{figure}[tbp] \label{f3}
\plotone{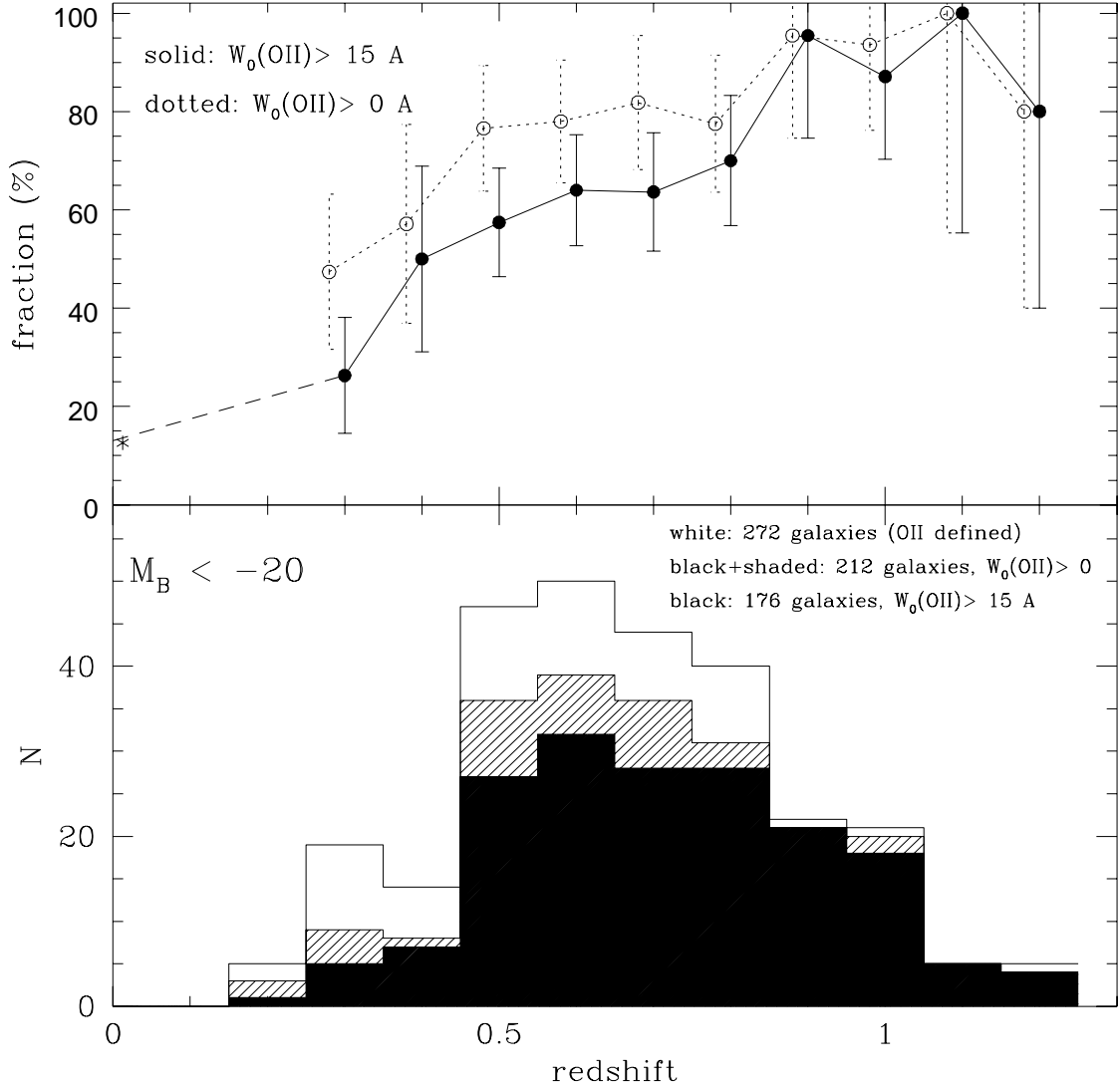}
\caption[]{{\it bottom}: redshift histogram for the 272 $M_{B} < -20$
galaxies (sample C); the black+shaded area represents the $W_{0}(OII)>$0 
galaxies (sample D) and the dark area the objects with significant
 star formation
($W_{0}(OII)>$ 15\AA). {\it (top)}: fraction of objects with $W_{0}(OII)>$
15\AA\ (solid line) and with $W_{0}(OII)>$ 0\AA; they increase by large
factors especially when compared to similar galaxies at low redshifts
(point at z=0 from Vettolani et al. 1996 and private communication).}
\end{figure}

\subsubsection{Selection effects against quiescent objects at high
redshift}

Crampton et al. (1995) have extensively studied the photometric
properties of the unidentified CFRS objects (15\% of the sample). They
concluded that at least half of the unidentified objects are likely to
be galaxies at the higher redshifts. This is supported by a simple
analysis of the [OII] 3727 emission line properties of the CFRS
galaxies.  Figure 3 shows the histogram and the fraction of
emission-line galaxies for $M_{B} < -20$. The fraction of emission-line
galaxies (i.e.  detectable equivalent width of [OII] 3727, $W_{0}(OII)
>$ 0) smoothly increases with redshift reaching a value of $\sim$75\%
from z = 0.5 to z = 0.8.  We believe that the virtual absence of
quiescent galaxies ($W_{0}(OII) = 0$) at z $> $0.8 is related to the
combination of at least two selection biases: (i) faint objects without
emission are more difficult to identify (see CFRS V); (ii) red and
early type galaxies at high-z are very faint at $\sim$ 6000\AA\ (peak
of transmission for our spectroscopy), which renders difficult redshift
measurements. For example, at 0.75 $ < z <$ 0.85, there is still a
significant fraction (9 among 25 galaxies) of quiescent galaxies in the
very luminous bin (sample C, $M_{B}<$-21), but there are none in the
less luminous bin (sample C, $-21 < M_{B} < -20$) at z$>$ 0.75 (0 among
24 galaxies). At these redshifts the latter bin is populated by
galaxies with $I_{AB}$ very close to our spectroscopic limit (22.5). If
the fraction of emission-line galaxies is assumed to remain constant at
75\% in the highest redshift ranges, the number of quiescent high-z
objects missed by our spectroscopy can be estimated.  There are only 3
quiescent galaxies among the 63 at z$> $0.85.  To maintain the 75\%
value beyond z = 0.75, we estimate that up to 16 objects (out of 410)
could have been missed because they were faint high-z galaxies without
emission lines. This implies that nearly a third ($\sim$ 25 galaxies)
of the CFRS failure rate is related to quiescent galaxies at high
redshift, in agreement with the discussion in CFRS V.

\section{EVOLUTION OF GROSS SPECTRAL PROPERTIES WITH REDSHIFT}

\subsection{[OII] 3727 emission line}

\subsubsection{ [OII] 3727 equivalent widths and luminosities}

Figure 3 shows that the fraction of bright emission-line galaxies
($W_{0}(OII) >$ 0 and $M_{B} <$-20) smoothly increases with redshift,
to $\sim$75\% at $z > 0.5$. Note that 62\% of the 141 $M_{B} <$ -20
galaxies with $ 0.45 < z < 0.75$ show substantial [OII] 3727 equivalent
widths ($W_{0}(OII) > $15\AA), which can be compared to 34\% of similar
objects among the 38 $M_{B} <$ -20 galaxies with $0.15 < z < 0.45$. The
redshift increase of the fraction of galaxies with significant line
emission is found to be highly significant even after accounting for
the quiescent galaxies which could have been missed by our spectroscopy
in the $ 0.45 < z < 0.75$ redshift interval. Probabilities that such an
increase occurs in a random distribution are as low as $10^{-7}$ (two
populations, Student t test, t=10 for $\nu$=228), even after assuming
that 50 quiescent galaxies with $M_{B}<$-20 have been missed in the $
0.45 < z < 0.75$ redshift interval (which is a crude overestimation,
see CFRS V). This trend is also supported by a comparison with what is
found in local samples of bright galaxies. Indeed, Vettolani et al.
(1996, and private communication) found that, in the ESP sample, 13\%
of the $M_{B} <$-20 galaxies show [OII] 3727 emission line with
$W_{0}(OII) > $15\AA. Recall that the ESP sample is well suited for a
comparison with the CFRS, since it includes the spectra of $\sim$ 4500
$B_{J} <$19.4 objects, regardless of their surface brigthness profiles,
and with a spectroscopic success rate of 95\%. From the DARS sample,
Peterson et al (1986) found similar values (17\% of the galaxies with
$W_{0}(OII) > $15\AA, with $>$ 70\% of the DARS galaxies having $M_{B}
<$-20). We are aware that possible biases related to aperture effects
can affect the comparison of the CFRS sample to local samples. These
effects are likely complex and would require a knowledge of the spatial
distribution of $W_{0}(OII)$ in a complete sample of local galaxies.
However we believe that they could not affect the major conclusion of
this section, i.e. there is a considerable increase of the fraction of
bright emission line galaxies from z=0 to z$>$0.5.  Firstly,
$W_{0}(OII)$ is a relative quantity, i.e. the ratio of the nebular
emission of the gas (related to the strength of the ionisation by very
hot stars) to the continuum emission at 3727A which is also related to
hot stars.  Secondly, the peak in the Vettolani redshift histogram is
at z=0.1, and their corresponding 2".4 diameter fiber represents 6 kpc
at z=0.1, i.e. it samples a reasonably large fraction of the galactic
disks. Thirdly, there is a good agreement between the two values found
in two differently selected samples (DARS and ESP have limiting
magnitude of $B_{J} =$16.5-17, and $B_{J} =$ 19.4, respectively), which
suggests that aperture effects cannot be dramatic when comparing the
properties of $W_{0}(OII)$ in different samples.  Our conclusion is
that between z = 0 and z = 0.5, there is a significant increase in the
fraction of luminous galaxies which show signs of star formation. This
mirrors the change in the restframe $(U-V)_0$--$M_B$ diagram shown in
CFRS VI (their Fig 5).  Beyond z=0.5 the fraction of galaxies with
$W_{0}(OII)>$ 15\AA\ is reaching a plateau with a nominal value from
55\% to 65\%, and the further increase beyond z=0.8 is likely related
to the difficulty in identifying quiescent galaxies at high redshift
(see section 2.4.2).

\begin{figure}[tbp] \label{f4}
\plotone{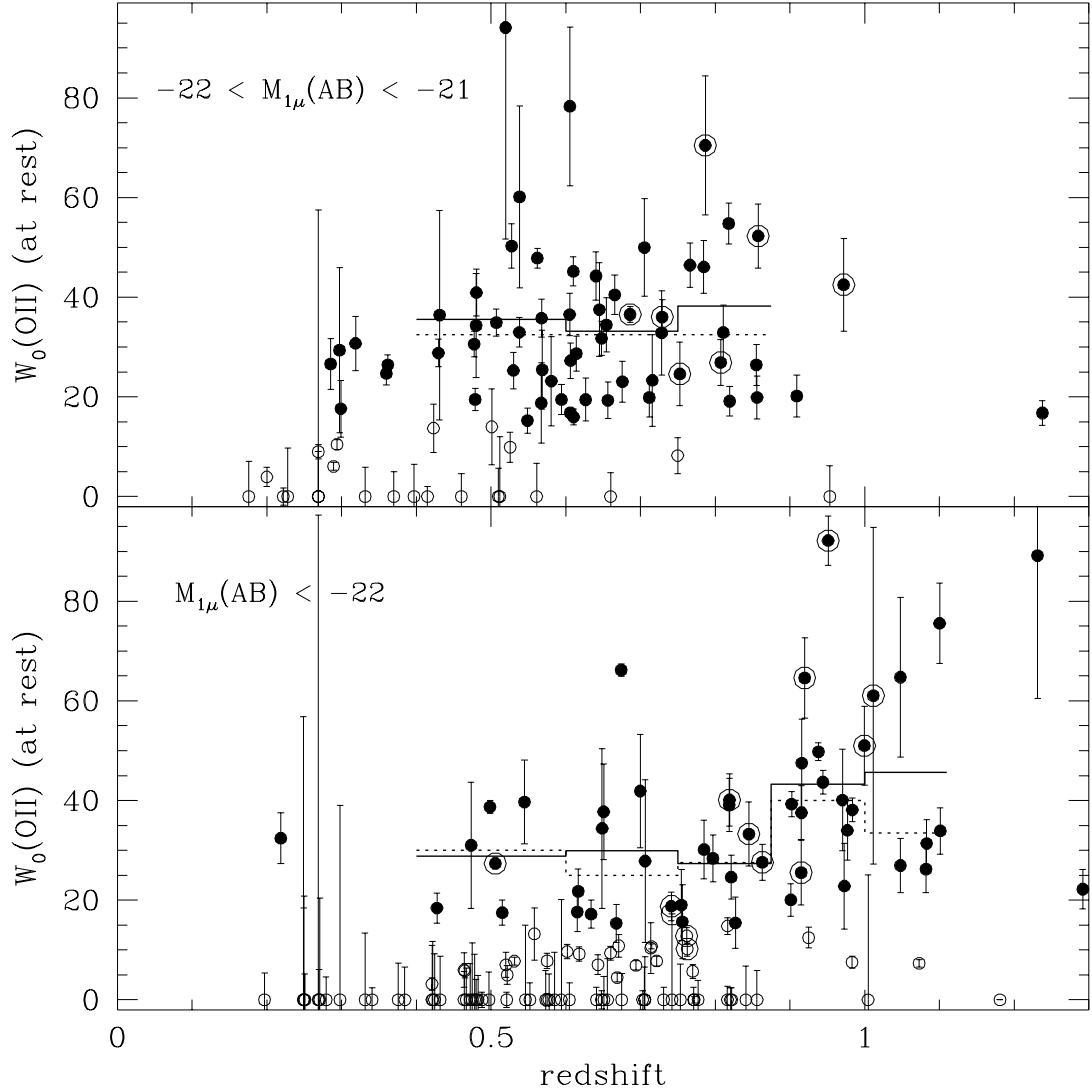}
\caption[]{Evolution of $W_{0}(OII)$ with redshift in two 
luminosity bins. Filled circles represent galaxies with significant emission
lines ($W_{0}(OII)>$ 15\AA). The redshift evolution for the latter is
shown by the solid and dotted lines, which correspond to the mean and median values,
respectively. These quantities have been calculated in redshift bins
of constant covolume for $q_{0}=$0.5 (see text and Table 3). In 
both luminosity bins, $W_{0}(OII)$ 
show no significant redshift increase from z = 0.4 to z = 0.875. In the
highest luminosity bin (bottom, $M_{1\mu} < -22$), $W_{0}(OII)$ mean and median values
 increase by
a factor 1.5 at higher redshift (z$ > $0.87). Circled dots distinguish 
D(4000)-deficient objects (only valid below z = 1).}
\end{figure}

In the sample C ($M_{B} < -20$ and [OII] 3727 in emission), the
increase of $W_{0}(OII)$ is rather modest (factor 1.4) between z = 0.4
(average $W_{0}(OII)$= 21.5\AA) and z = 0.8 (average $W_{0}(OII)$=
29.6\AA).  For comparison, in the DARS sample (Peterson et al. 1986)
the average $W_{0}(OII)$ of the emission-line galaxies is 20\AA.  In
order to quantify the redshift evolution of $W_{0}(OII)$, we have
restricted our sample to the objects showing significant emission lines
($W_{0}(OII)>$ 15\AA).  One can argue that these galaxies up to z=1 and
slightly beyond cannot have been missed in our spectroscopic sample (at
z$\sim$ 1 they would show an equivalent width of 30\AA\ in the observer
frame). Furthermore, we have subdivided the sample in 5 redshift bins
(see Table 3), the motivation of the choice of the redshift ranges
being to have a fair equipartition of the covolume from z=0.4 to z=1.11
in 5 equal parts (assuming $q_{0}=$0.5). From Table 3, one can
investigate the redshift evolution of the comoving number density of
galaxies with significant emission lines, and of the median and average
values of $W_{0}(OII)$.  We have also divided the sample of
emission-line galaxies according to their absolute magnitude, at B and
at 1$\mu$ wavelengths, respectively.  In both sub-samples ($M_{B} <
-21$ and $-21 < M_{B} < -20$ galaxies, respectively), the increase of
the median and of the mean values of $W_{0}(OII)$, is small and not
significant.  Figure 4 shows the redshift evolution of $W_{0}(OII)$ for
galaxies in the two $1\mu$ luminosity bins. In the sub-sample of
luminous galaxies at $1\mu$ ($M_{1\mu}<-22$), the mean and median
values of $W_{0}(OII)$ are constant from z=0.4 to z=0.875, and suddenly
increase in the two higher redshift bins. A simple statistical test
shows that the two populations (at $0.4<z<0.875$ and at $0.875<z<1.11$,
respectively) cannot be drawn from the same population (2 populations,
Student t test, t=3.9 for $\nu$=46, P=0.9998), relatively to their
$W_{0}(OII)$ properties. It can be also interpreted as an increase of
the actual scatter of $W_{0}(OII)$ values (of $M_{1\mu}<$ -22 galaxies)
with the redshift, as it can be seen from Table 3.

\begin{deluxetable}{crcccr}
\tablenum{3}
\tablewidth{0pt}
\tablecaption{Comoving statistical properties of $W_{0}(OII)>$ 15A galaxies}
\tablehead{ 
\colhead{M range} & \colhead{z range} & \colhead{N objects} & \colhead{median $W_{0}(OII)$} & 
\colhead{mean $W_{0}(OII)$} & \colhead{$\sigma$}}
\startdata
$M_{B} < -21$ & 0.40-0.60 & 10 & 32.5 & 33 & 13.23 \nl
$M_{B} < -21$ & 0.60-0.75 & 12 & 31 & 31.9 & 14.01 \nl
$M_{B} < -21$ & 0.75-0.875 & 11 & 34 & 39.3 & 19.61 \nl 
$M_{B} < -21$ & 0.875-1.00 & 21 & 38 & 40.9 & 18.37 \nl
$M_{B} < -21$ & 1.00-1.11 & 10 & 32.5 & 41.4 & 18.37 \nl
$-21<M_{B}<-20$ & 0.40-0.60 & 21 & 32.5 & 37 & 20.14 \nl
$-21<M_{B}<-20$ & 0.60-0.75 & 31 & 32.5 & 32.2 & 13.74 \nl 
$-21<M_{B}<-20$ & 0.75-0.875 & 18 & 32 & 36.3 & 16.53 \nl 
$-21<M_{B}<-20$ & 0.875-1.00 & 5 & 36 & 28.7 & 3.1 \nl
$M_{1\mu}<-22$ & 0.40-0.60 & 6 & 30 & 28.8 & 9.58 \nl
$M_{1\mu}<-22$ & 0.60-0.75 & 10 & 25 & 29.9 & 15.84 \nl
$M_{1\mu}<-22$ & 0.75-0.875 & 10 & 27.5 & 27.3 & 8.82 \nl
$M_{1\mu}<-22$ & 0.875-1.00 & 14 & 40 & 43.3 & 18.42 \nl
$M_{1\mu}<-22$ & 1.00-1.11 & 7 & 33.5 & 45.7 & 20.67 \nl
$-22<M_{1\mu}<-21$ & 0.40-0.60 & 19 & 32.5 & 35.5 & 18.38 \nl
$-22<M_{1\mu}<-21$ & 0.60-0.75 & 21 & 32.5 & 33.2 & 14.26 \nl
$-22<M_{1\mu}<-21$ & 0.75-0.875 & 11 & 32.5 & 38.2 & 16.8 \nl
$-22<M_{1\mu}<-21$ & 0.875-1.00 & 2 & 31 & 31.3 & - \nl
\enddata
\end{deluxetable}

\begin{figure}[tbp] \label{f5}
\plotone{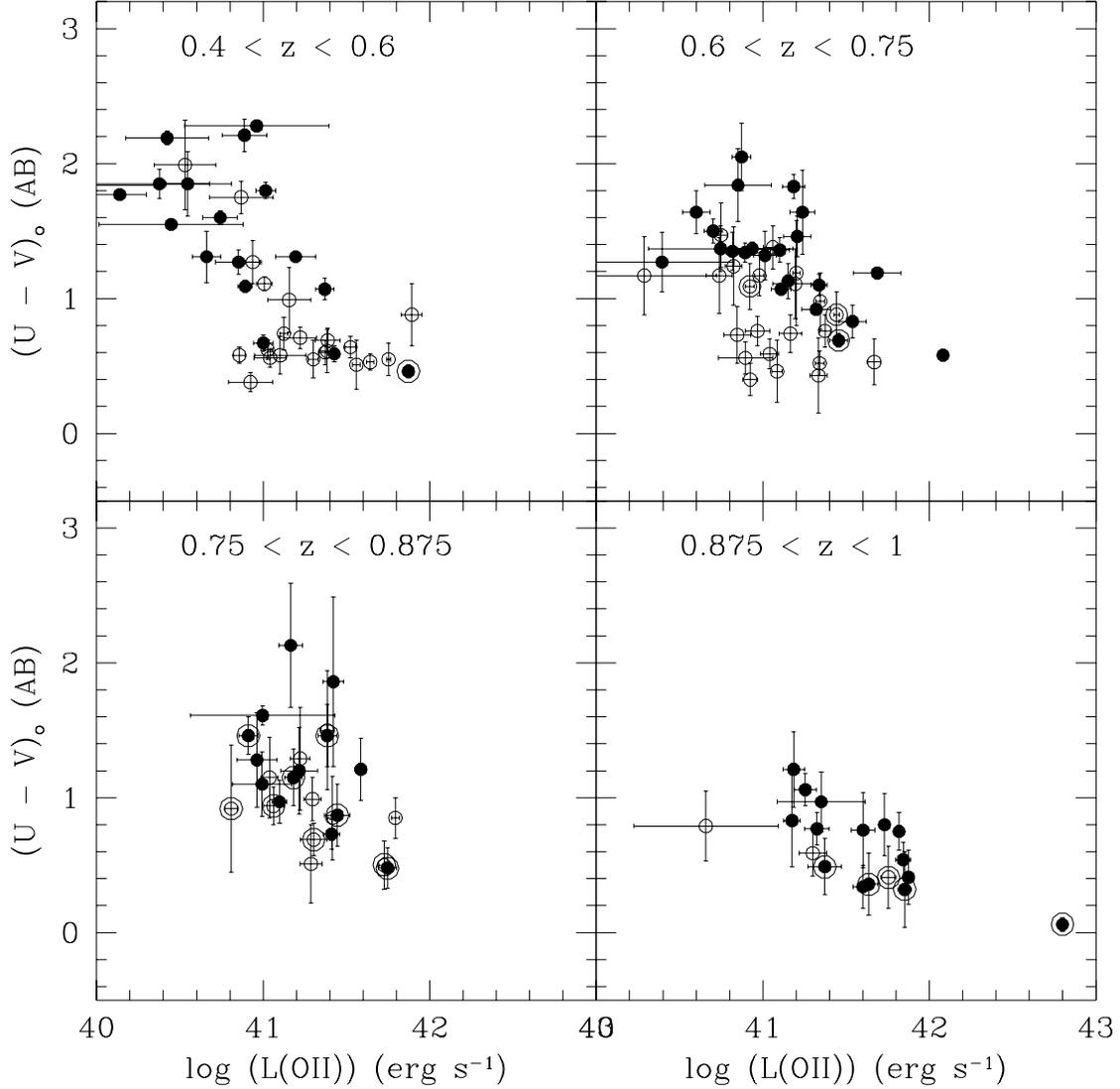}
\caption[]{Relation between rest frame $(U-V)_{AB}$ colors and [OII] 3727
luminosity  in 4 redshift bins, chosen to provide an equipartition of
the covolume from z = 0.4 to z = 1 ($q_{0}$=0.5). Filled circles represent
galaxies with substantial amounts of old stars ($M_{1\mu}< -22$), open
circles the rest of the sample ($-22 < M_{1\mu} < -21$) and circled dots
are the D(4000)-deficient galaxies.}
\end{figure}

Evolution of the most luminous galaxies at 1$\mu$ is corroborated by
examination of the relation between the rest frame $(U-V)_{AB}$ colors
and the [OII] 3727 luminosities (Figure 5), which are known to be
linked to the birthrate parameter (observed rate compared to the
average past rate of star formation) and to the observed rate of star
formation, respectively (see Kennicutt et al. 1994). The comoving
number density of luminous galaxies ($M_{1\mu} < -22$) shows no
increase in the 4 redshift bins, while with increasing redshift, they
apparently shift towards bluer $(U-V)_{AB}$ colors (median in the
highest redshift bin 1 mag bluer than in the lowest redshift bin) and
higher [OII] 3727 luminosities. This effect is followed by only a
modest shift of the rest frame $(B-1\mu)_{AB}$ color (0.4 mag for the
median), so the color shift is mainly related to UV light. This can be
interpreted as due to either an effective blueing of the most 1$\mu$
luminous objects accompanied by a large increase of the SFR, or to the
emergence at high redshift of a 1$\mu$ and [OII] 3727 luminous
population of galaxies which should fade away at lower redshift.  The
emission-line galaxies with the highest luminosities at 1$\mu$ show
significant changes with redshift in both $(U-V)_{AB}$ color and in
[OII] 3727 luminosity. These changes are similar to those found by
Cowie et al (1996) in the Hawaii deep fields (see their Figure 10).

\subsubsection{ [OII] 3727 luminosity comoving density}

The comoving luminosity density of [OII] 3727 has been computed in 4
redshift bins, following the $V_{max}$ formalism used to construct the
continuum luminosity densities in CFRS XIII. [OII] 3727 luminosities have
been calculated from the measured [OII] 3727 fluxes. Hence we have applied
aperture correction factors (with average values ranging from 1.4 to
1.6 in the 4 redshift bins) to account for the light missed by our
1\farcs75 slit. 

 Aperture corrections have been calculated in two different ways: (i)
we have computed the fraction of V broad band light which has been missed by
our spectroscopy (recall that for z$>$0.4 the V filter samples the UV light
at rest); (ii) we have estimated the continuum flux at 3727A for all galaxies
from their $V_{AB}$ magnitude and $(V-I)_{AB}$ color (see CFRS VI) and
have calculated the [OII] 3727 flux from $W_{0}(OII)$. Both methods lead to
similar correction factors, without strong variations from one redshift bin
to another. 
 
    The ``directly-observed'' comoving luminosity density
was first computed based on the $I_{AB} \le 22.5$ CFRS galaxies with
measured [OII] 3727 fluxes. Note that these estimates are unlikely to
be affected by our failure to obtain redshift identifications for
19$\%$ of the galaxies, since the latter are unlikely to have [OII]
3727 emission lines, but they will be lower limits since there are
galaxies fainter than the survey limit of $I_{AB} < 22.5$.  Statistical
uncertainties in these estimates have been calculated as in CFRS XIII.

\begin{figure}[tbp] \label{f6}
\plotone{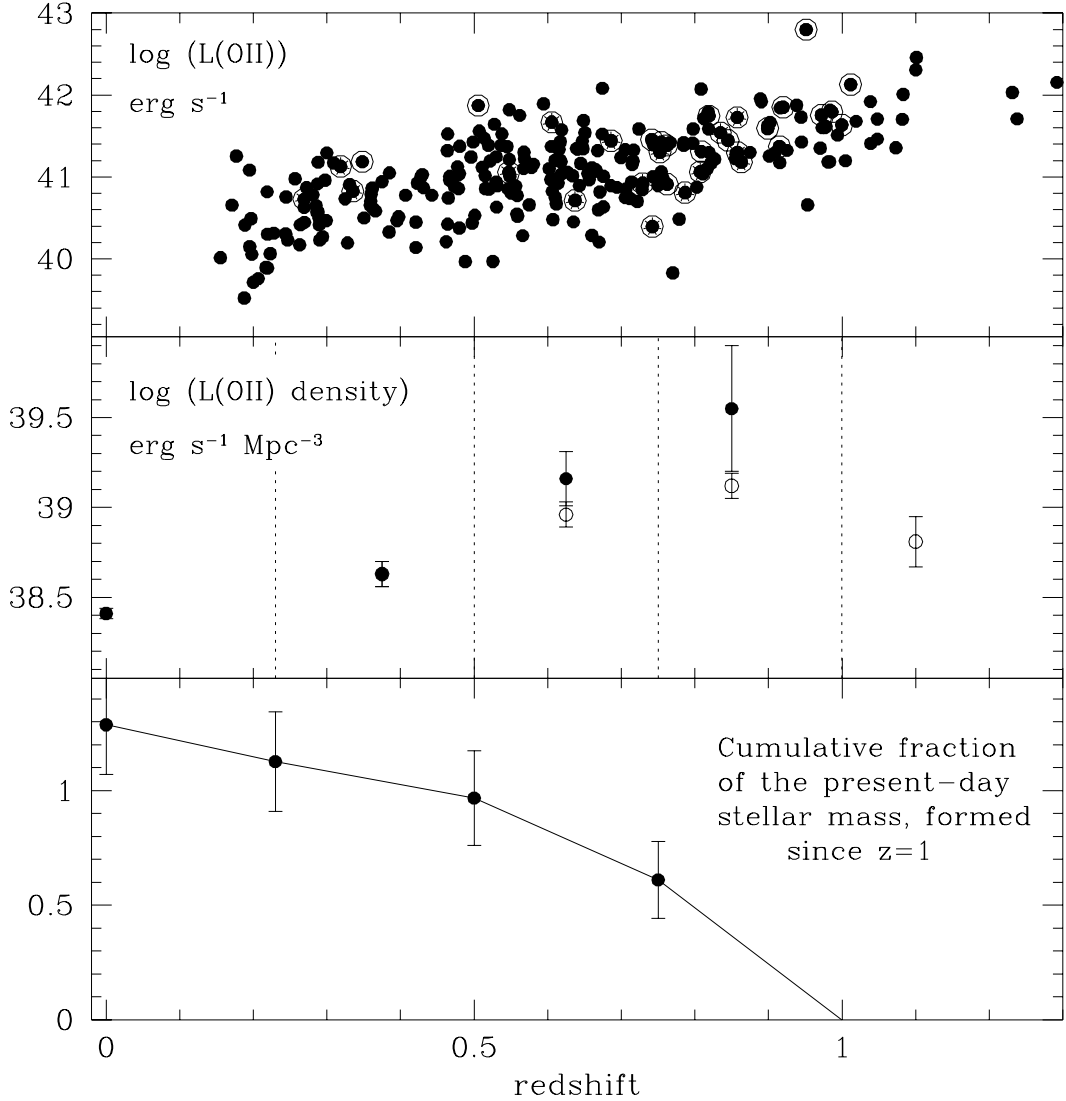}
\caption[]{{\it (top)}: [OII] 3727 luminosity against redshift (sample B).
{\it (middle)}: redshift evolution of the comoving density of the [OII]
3727 density; open dots represent the directly observed values and full
dots the estimated values assuming contributions of low luminosity
galaxies ($M_{B}<$-18.5) at all redshifts (see text and Table 4).  The
point at z = 0 is from Gallego et al. assuming the K92 relationship
between $[OII]$ 3727 and $H\alpha$ luminosities. Since Gallego et al.
integrate over all luminosities, the z = 0 value is an upper limit
relative to the values at higher z.  {\it (bottom)}: cumulative
fraction of the present day stellar mass which would be formed since z
= 1, assuming Kennicutt's SFR calibration; solid line  represents the
contribution of all galaxies with $M_{B} < -18.5$.}
\end{figure}

The contribution of the low luminosity galaxies undoubtedly lying below
the CFRS magnitude limit has been estimated as follows.  The low
redshift bin (0.23 $<$ z $<$ 0.5) includes $M_{B} < -18.5$ galaxies,
the intermediate redshift bin (0.5 $<$ z $<$ 0.75) $M_{B} < -19.5$
galaxies, and the high redshift bin (0.75 $<$ z $<$ 1)  $M_{B} < -20.5$
galaxies. Three scenarios have been considered to estimate the [OII] 3727
luminosity density produced by the ``missing'' low luminosity galaxies
in the intermediate and high redshift bins:  (i) strong evolution
similar to that of $M_{B} < -19.5$ or of $M_{B} < -20.5$ galaxies,
respectively; (ii) no evolution at $z \ge 0.35$; (iii) a mixed
evolution scenario, in which $M_{B} > -19.5$ galaxies would not evolve,
but $ -20.5 < M_{B} <-19.5$ galaxies would.  Figure 6 shows the average values
resulting from these three scenarios which all fall very close to those
from the mixed evolution scenario.  The error bars include the
differences among the three scenarios and become large for the high
redshift bin.  These values are given in Table 4. At very low redshifts, 
the luminosity density at z = 0
has been derived from Gallego et al. (1995) assuming the K92
relationship between [OII] 3727 and $H\alpha$ luminosities. 


\begin{deluxetable}{lllcl}
\tablenum{4}
\tablecaption{Comoving densities of [OII] 3727 luminosity ($H_{0}$=50)}
\tablewidth{0pt}
\tablehead{ 
\colhead{Redshift range}  & \colhead{directly observed} &
\colhead{} & \colhead{corrected for incompletness ($M_{B}<$-18.5 galaxies)} &\colhead{} }
\startdata
           &  $10^{39} erg s^{-1} Mpc^{-3}$ &             & $10^{39} erg s^{-1} Mpc^{-3}$ &\nl
           &  $q_{0}$=0.5  & $q_{0}$=0.1                   & $q_{0}$=0.5 & $q_{0}$=0.1 \nl

0.25-0.50 & 0.43 $\pm$0.07 & 0.38 $\pm$0.06 &      0.43 $\pm$0.07 & 0.38 $\pm$0.06 \nl
0.50-0.75 & 0.91 $\pm$0.15 & 0.77 $\pm$0.12 &      1.44 $\pm$0.42 & 0.86 $\pm$0.34\nl  
0.75-1.00 & 1.32 $\pm$0.23 & 1.08 $\pm$0.18 &      3.61 $\pm$2.10 & 2.36 $\pm$ 1.3\nl  
1.00-1.25 & 0.65 $\pm$0.11 & 0.5 $\pm$0.09  \nl 
\enddata
\end{deluxetable}

The [OII] 3727 luminosity density of field galaxies 
( $M_{B} <$-18.5) does not
appear to increase much from z = 0 to z = 0.375 (by a factor 1.6, over
an elapsed time equal to 28\% of the age of the Universe), but
increases by a large factor (8.4 $\pm$3.5) between z = 0.375 and z =
0.85 (elapsed interval 22\% of the age of the Universe).  The latter
increase is even larger than the evolution of the luminosity density
found at 2800\AA\ (factor 4.3$\pm$0.9, CFRS XIII) although the overall
increase over the whole redshift range $0 < z < 1$ is comparable.
 These results are quantitatively and qualitatively independent of 
$q_{0}$, given the error bars (see Table 4 and compare the values 
for $q_{0}$=0.5 and $q_{0}$=0.1, respectively).
 The comoving densities shown in Figure 6 (middle panel) are systematically
higher than those of Cowie et al. (1995), by factors of about 5, for
reasons we do not understand (the individual luminosities agree well,
statistically). The number quoted by Cowie et al. (1995) at z = 0.375
(6$\times10^{37}$ erg s$^{-1}$Mpc$^{-3}$) is much lower than the
derivation from Gallego et al. (1995) locally (2.6$\times10^{38}$ erg
s$^{-1}$Mpc$^{-3}$), which is not realistic. Our value at z = 0.375 is
not much higher than that of Gallego et al.  (Figure 6), which is
consistent with the modest evolution in the luminosity function of blue
galaxies and in disk surface brightness from z = 0 to z = 0.375 (CFRS
VI and CFRS IX).

\subsection{Continuum indices}

\subsubsection{Continuum indices versus redshift and duration of the star formation}

\begin{figure}[tbp] \label{f5}
\plotone{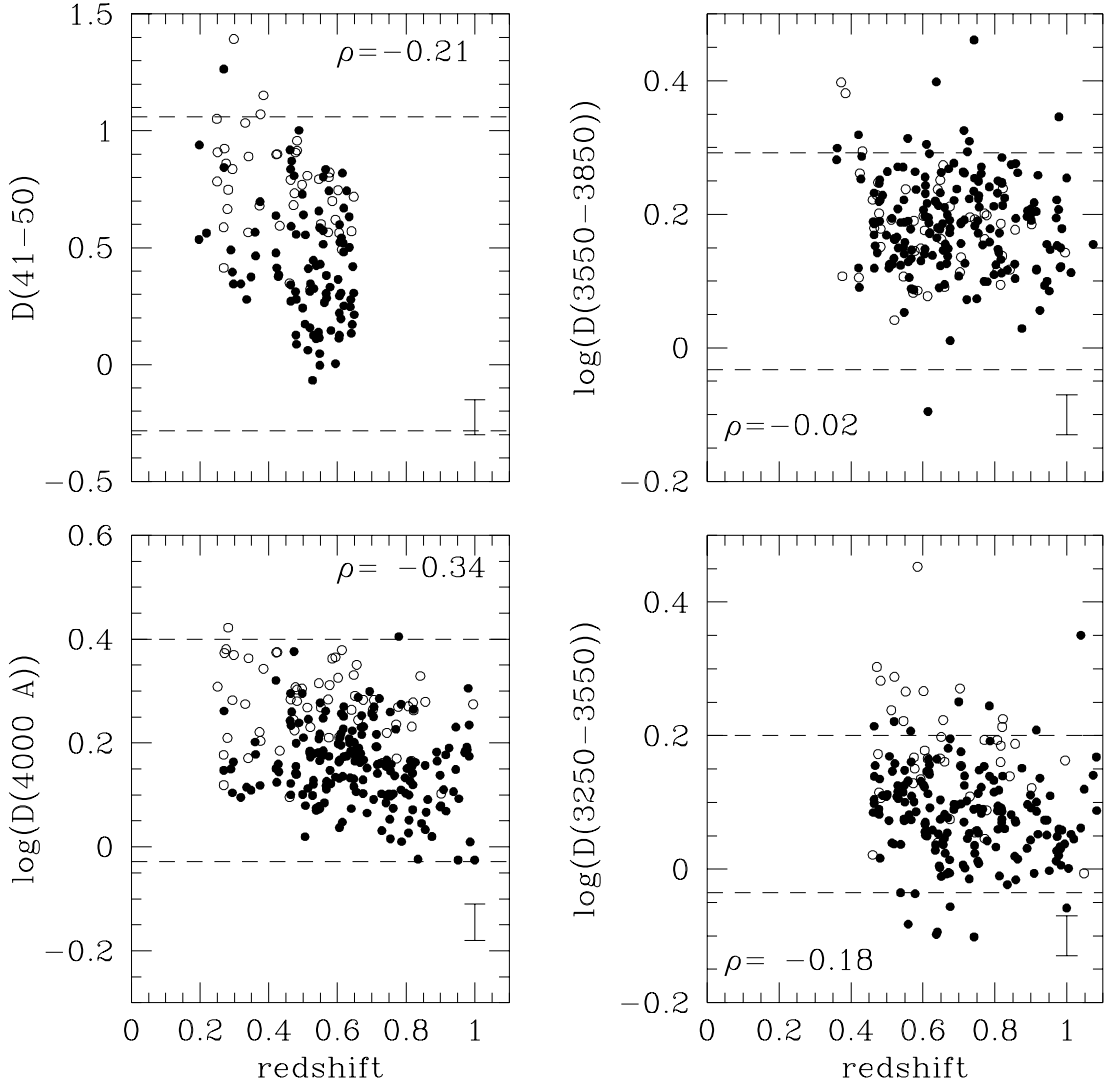}
\caption[]{ Evolution of four color indices with redshift for
galaxies with $M_{B}< -20$ (sample C); the full dots represent emission-line
galaxies, open dots quiescent galaxies (with $W_{0}(OII)$= 0). 
Correlation parameter are given in each panel. Dashed lines show 
the limits of the color indices from BC95. Typical error bars are
shown in the bottom left in each panel.}
\end{figure}

\begin{figure}[tbp] \label{f5}
\plotone{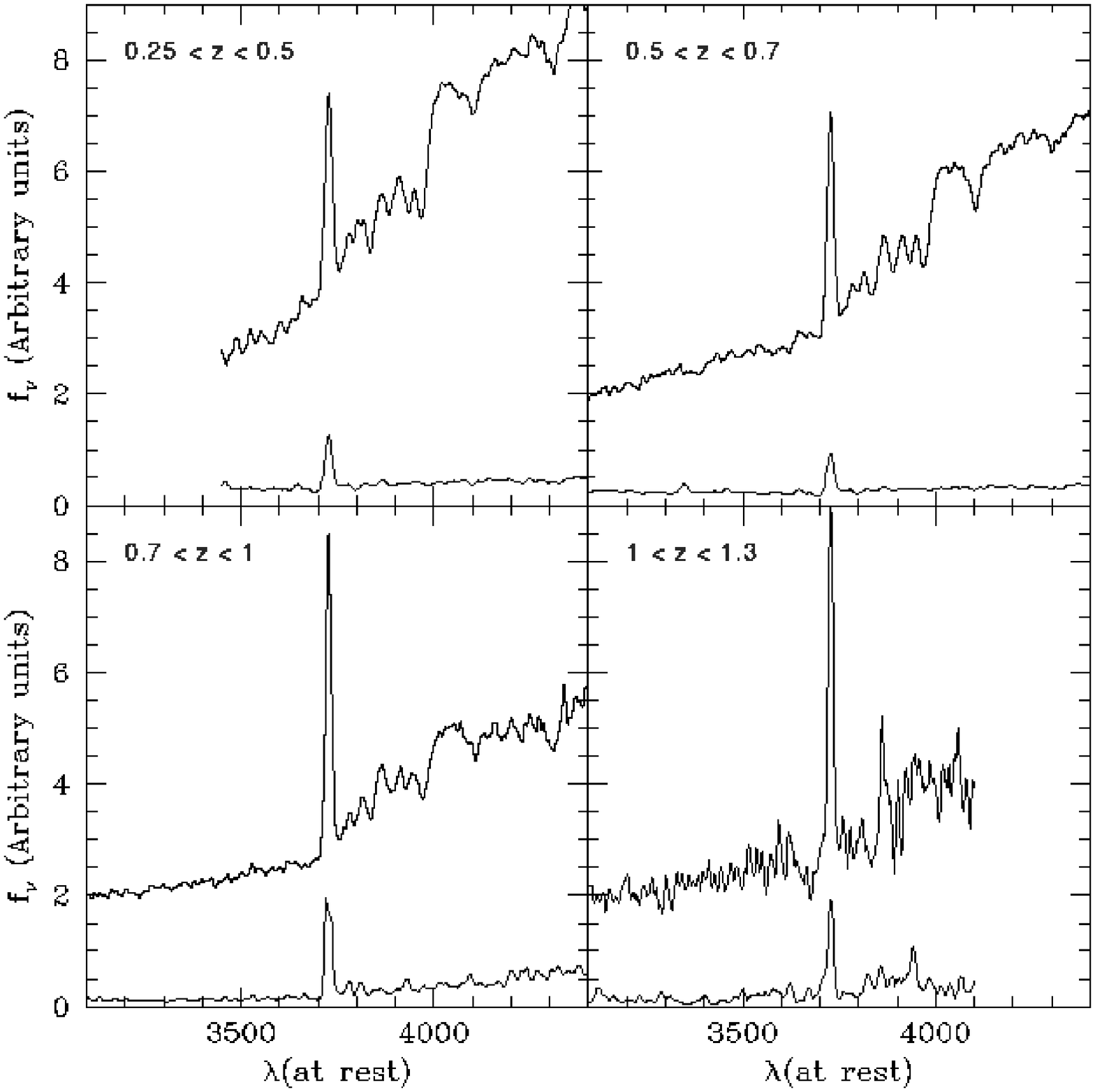}
\caption[]{Average spectra of $M_{B}< -20$ emission-line galaxies (sample D)
 in four redshift bins. Average has been computed by scaling each individual
spectrum in the 3500-4500\AA\ wavelength range (at rest). Residuals 
(1$\sigma$ deviation) are shown in the bottom of each panel.}

\end{figure}

Some of the continuum indices show a modest decrease or ``blueing" with
redshift (Figure 7) in the two luminosity bins. The D(4000) decrease is
statistically significant ($\rho$ from 0.3 to 0.4). On average, t
he D(41-50) index decreases
by $\sim$ 0.4 magnitude between z = 0.4 and 0.6, the D(4000) index by
$\sim$ 0.3 magnitude between z = 0.5 and 1. At lower wavelengths, the
UV index D(3250-3550) decreases by only $\sim$ 0.1 magnitude (not
significant) between z = 0.5 and z = 1  and the Balmer index
D(3550-3850) shows no apparent decrease with redshift.   The relatively
modest color evolution found, especially at the UV end of the spectra,
indicates that a dramatic increase in the fraction of extremely blue
and young galaxies (e.g., starbursts or blue compact galaxies with
$(U-V)_{AB}<$0) in the past is unlikely (see for example Figure 2). The
data suggest more modest evolution involving sustained star formation.

Figure 8 shows average spectra of emission-line galaxies in 4 redshift
bins. One can see that the increase of $W_{0}(OII)$ is accompanied by a
decrease of the 4000\AA\ break with the redshift, while relatively
large Balmer breaks are found at all redshifts.  Large values for the
latter are unavoidably accompanied by large values for the Balmer
absorption lines, for any plausible stellar population. In the average
spectra, typical values for $W_{0}(H\delta)$ are in the range between 3
and 4\AA. We have empirically set a limiting value of log D(3550-3850)
= 0.2 (which corresponds to equivalent widths for Balmer lines of
5-7\AA) to distinguish emission-line objects having a significant A
star population from the others. Emission-line galaxies with Balmer
indices higher than this value have experienced a significant burst
several tenths of Gyr ago, while they were still forming stars at the
time their observed light was emitted. Assuming that the burst duration
is at least as large as the burst age, the BC93 and BC95 codes provide
an estimate of the elapsed time since the beginning of the significant
burst. We find that log D(3550-3850) = 0.2 implies $T_{elapsed}>$ 0.6
Gyr assuming a constant star formation rate
( the accuracy in the latter value is better than 0.05 Gyr after
testing a complete range of time scale for the burst). 41\% of the 170
emission-line galaxies ($M_{B}< -20$, 0.45 $<z<$ 1) show a substantial
A star population according to the above criterion. This fraction is
found to be the same at all redshifts (Figure 7) and is unlikely to be
affected by the presence of extinction or by a mixture of stellar
populations.  The fact that a large fraction of the galaxy population
shows emission lines, and also has a significant A star population,
suggests that most of the field galaxies with emission lines, at
redshift ranging from z = 0.4 to z = 1, are experiencing long-duration
events of star formation (at least 0.6 Gyr). This is also consistent
with the relatively small number of post-starburst galaxies (section
8.2) in the sample, which represent only 11\% of the objects having a
significant population of A stars. Recall that a much larger fraction
of post-starburst galaxies is found in rich distant clusters,
interpreted as being related to short-duration bursts ($\sim$ 0.1 Gyr, Barger
 et al. 1996). Latter analysis is similar than our, and based on the 
relative strengths of the $H\delta$ equivalent width and on $B-R$ color
 indices of cluster galaxies.

\subsubsection{Continuum indices versus [OII] 3727}

\begin{figure}[tbp] \label{f5}
\plotone{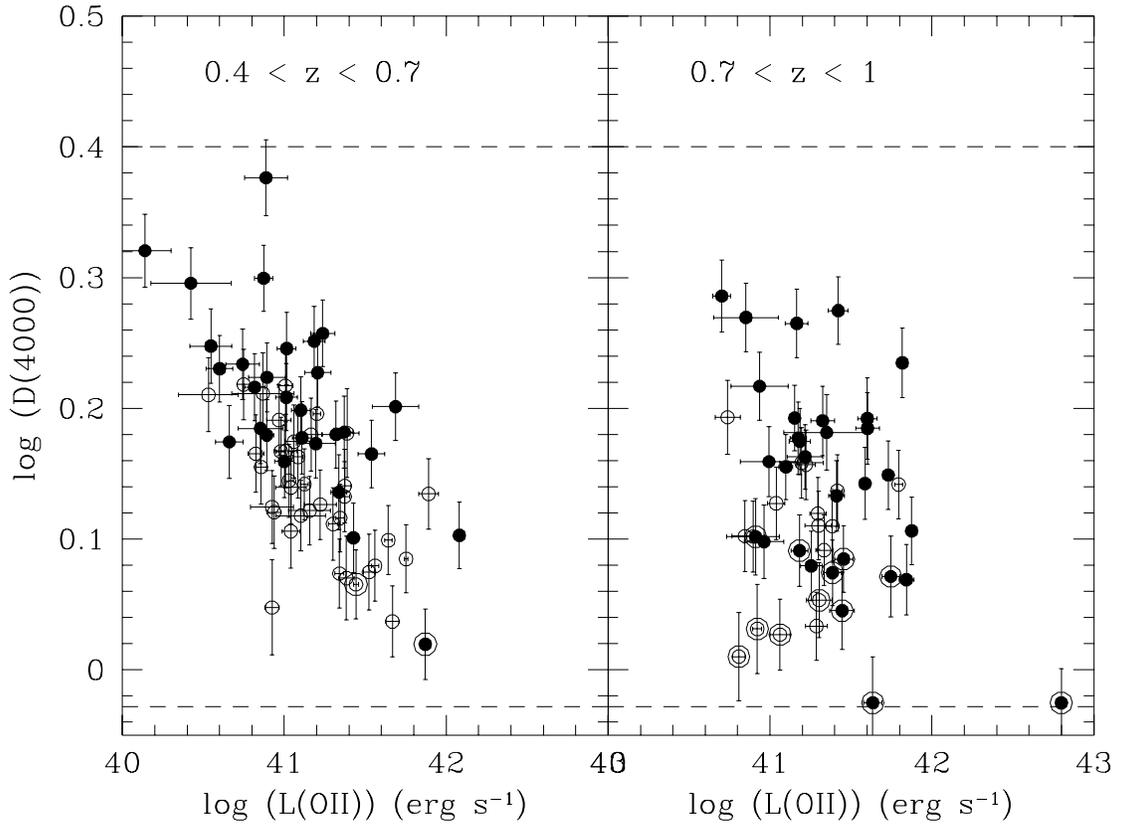}
\caption[]{Relation between D(4000) index and [OII] 3727 luminosity 
in two redshift bins. Full dots represents galaxies with
substantial amounts of old stars ($M_{1\mu}< -22$); open dots the rest
of the sample ($-22 < M_{1\mu} < -21$). The circled dots are the
D(4000)-deficient galaxies. The dashed lines show the limit on the 
D(4000) index from the BC95 model.}

\end{figure}
\begin{figure}[tbp] \label{f5}
\plotone{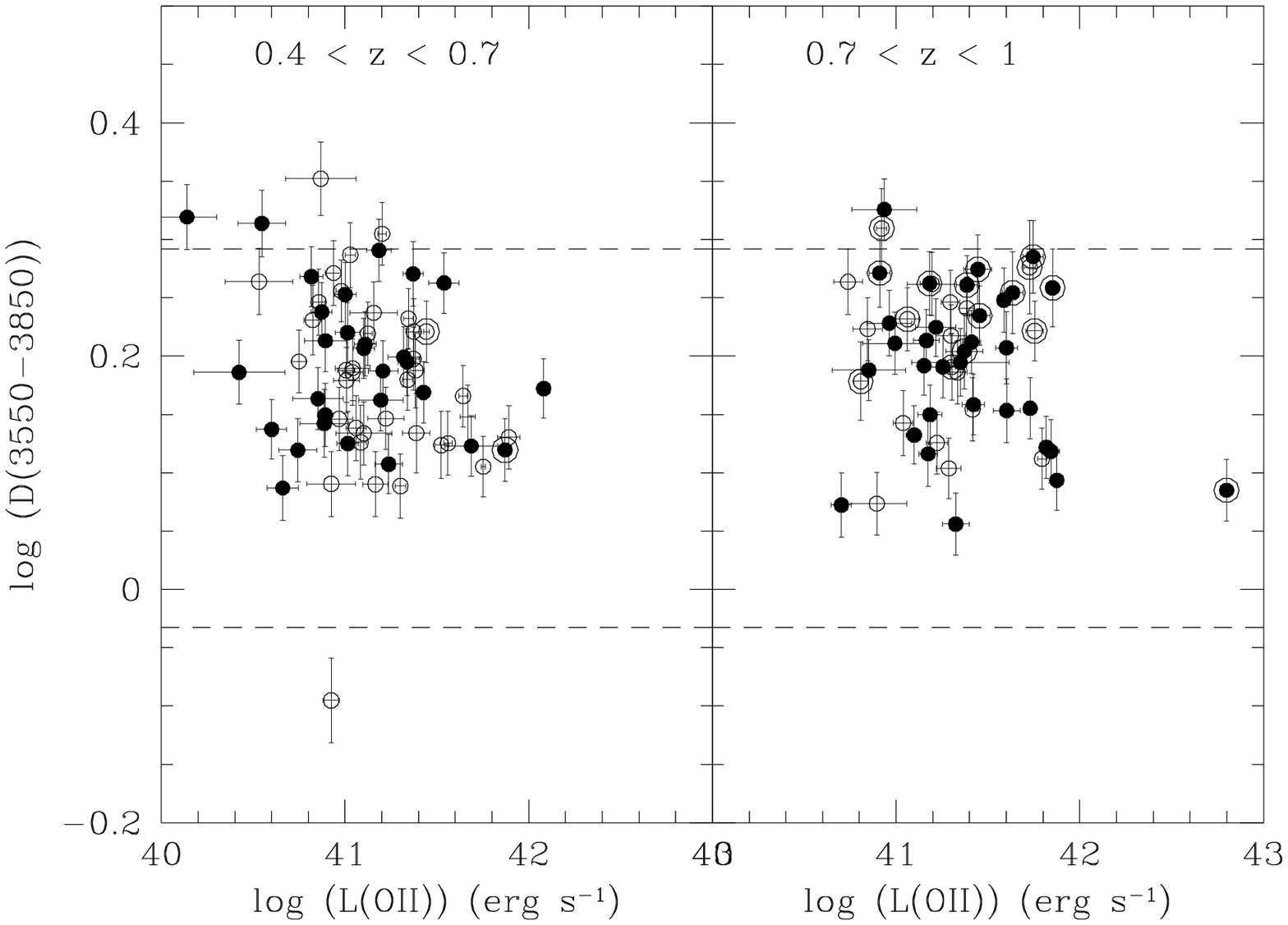}
\caption[]{Relation between Balmer index and [OII] 3727
luminosity in two redshift bins. Full dots represents
galaxies with substantial amounts of old stars ($M_{1\mu}<  -22$); open
dots the rest of the sample ($-22 < M_{1\mu} < -21$). The circled dots
are the D(4000)-deficient galaxies. The dashed lines show the limit on the 
Balmer index from the BC95 model.}

\end{figure}

The relationships between the [OII] 3727 emission line and  continuum
indices are shown in Figures 9 and 10. As expected, the bluest
continuum indices are associated with the strongest emission lines, and
vice versa.  At intermediate redshifts (0.4 $<$ z $<$ 0.7) the D(4000)
index correlates with the [OII] 3727 luminosity while at higher redshift
there is a strong increase in the dispersion of the relation. Interestingly,
 the relationship between D(4000) indices and $W_{0}(OII)$ (which are both
relative quantities) shows the same trend (strong correlation in the
moderate redshift bin, no correlation in the high redshift bin).  At
all redshifts and for every [OII] 3727 luminosity (or $W_{0}(OII)$), there
is a substantial fraction of galaxies having large Balmer indices,
indicative of the presence of a population of A stars.  More generally,
there is a difference in the behavior of ``red" indices, D(4000) and
D(3550-4150), relative to the UV color indices, D(3250-3550) and
D(3550-3850).  The ``red" indices show a net decrease with [OII] 3727
luminosity and with redshift in contrast to the UV indices which stay
roughly constant. The latter is found to be largely independent of OII
properties at all redshifts.  Galaxies with high [OII] 3727 luminosities have
UV color indices much redder than an extremely young burst (such as a
blue compact dwarf, see the lower limit drawn in Figure 10). We find
essentially no objects that would be comparable to the extreme blue
compact galaxies (see Izotov et al.  1995), i.e. with $W_{0}(OII)$ larger
than 100\AA\ and extremely blue color indices.

\subsection{Summary: Gross evolutionary properties from z = 0.4 to 1}

The most obvious sign of spectral evolution of field galaxies is the
huge increase in the fraction of galaxies showing significant star
formation ($W_{0}(OII) >$ 15\AA), from 13\% locally to 50\% or more at z
$>$ 0.5. Beyond z = 0.5 there is a certain ``saturation" in the
evolution of the population of the less luminous galaxies ($-22<
M_{1\mu} < -21$) which show no further evolution of their broad
band-colors or [OII] 3727 properties (Figures 4, 5).

At  $z > 0.5$, the decrease of the rest frame $(U-V)_{AB}$ color with
redshift is mostly associated with galaxies with large near-infrared
luminosities ($M_{1\mu} < -22$, Figure 5). The ``blueing" of galaxies
in this category with high redshift is then mostly due to a change in
their spectral energy distributions from 3850\AA\ to 5500\AA, without
color changes below 3850\AA. In other words, beyond z = 0.5, the colors
of the population of luminous galaxies with $M_{1\mu} < -22$ shift
closer to those of less luminous galaxies. In this redshift range, the
strong redshift evolution of the [OII] 3727 luminosity density is mainly
 due to the increasing contribution of galaxies that are luminous at $1\mu$.

The large values reached by the comoving [OII] 3727 luminosity density at z
= 1 suggest that a significant fraction of the stars have been formed
since z = 1. This is supported by the fact that most of the
emission-line galaxies (which represent 75\% of the z$>$0.7 galaxies)
are probably experiencing relatively long ($>$ 0.6 Gyr) epochs of star
formation. These events are accompanied by a consequent brightening of the
disk surface brightness (1 mag from z = 0.3 to z = 0.8, CFRS IX).
All of these are consistent with the fact that, on average, CFRS galaxies
 with significant star formation ($W_{0}(OII) >$ 15\AA) have restframe
$B-1\mu$ colors that are 0.6 mag bluer than the rest of the sample.
Assuming an exponentially decreasing SFR ($\tau$$\sim$0.8 models), a
brightening of $\sim$ 1 mag (at $B$ wavelength which corresponds to a 
0.6 mag blueing of the $B-1\mu$ color index) is produced if
 $\sim$ 10\% of the galaxy
mass is formed during the corresponding period of star formation.
Since beyond z = 0.5, $>$50\% of the galaxies have significant star
formation, it is possible that star formation has been sustained in
many galaxies as long as several Gyrs, leading to the formation of a
substantial fraction of their masses between z = 1 and z = 0.5 (elapsed
time $\sim$2.3 Gyr).

\section{EMISSION LINE PROPERTIES OF z $<$ 0.7 GALAXIES}

In this section we look in detail at the line ratios of CFRS
emission-line galaxies, limiting our attention to $z < 0.7$ since it is
only in this redshift range that [OII] 3727, H$\beta$ and [OIII] 5007
can be observed in the CFRS spectral range.

\begin{figure}[tbp] \label{f5}
\plotone{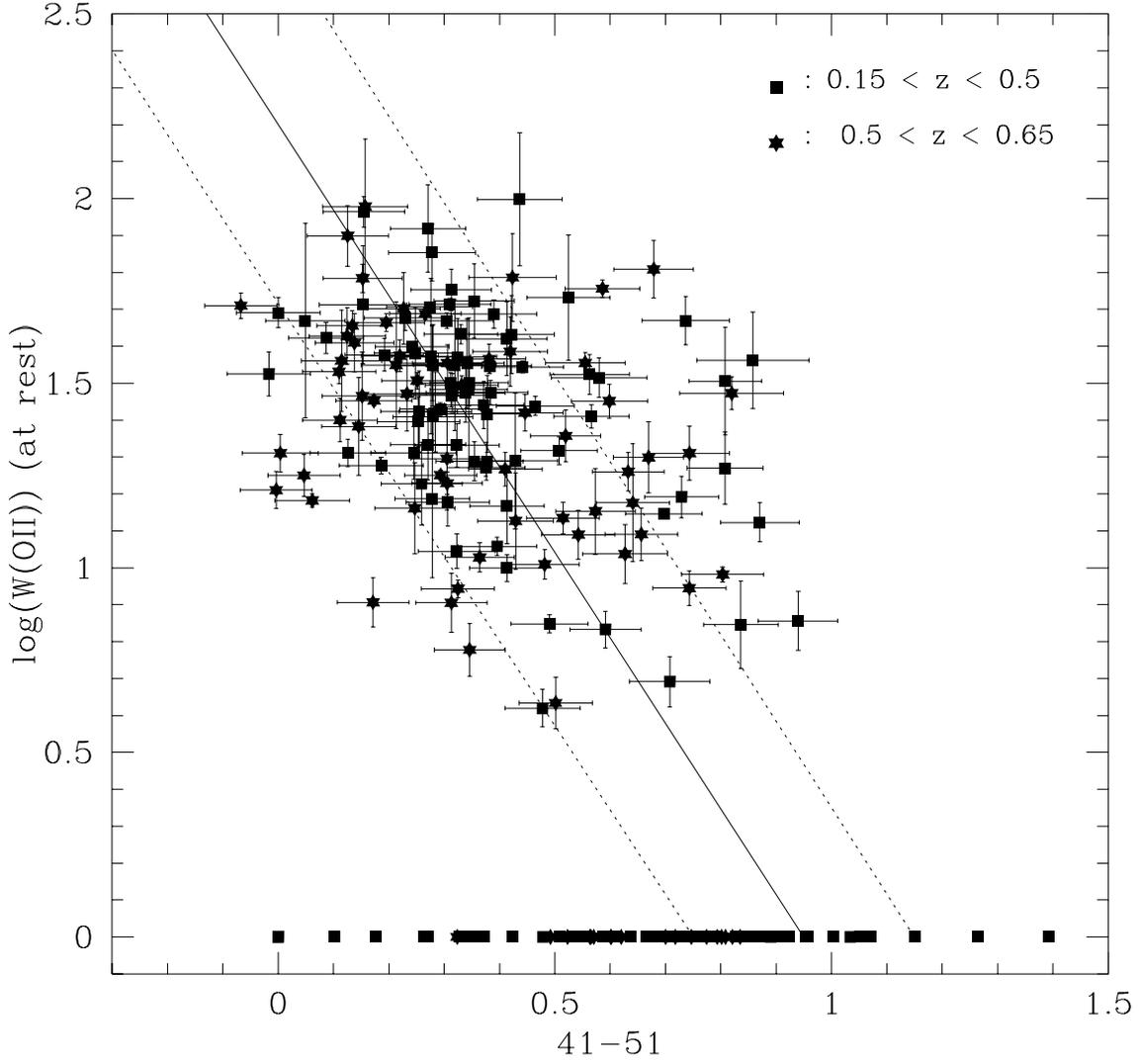}
\caption[]{ Relation between the [OII] 3727 equivalent width and the D(41-50)
continuum index for 139 z $<$ 0.5 galaxies (squares symbols) and 101
0.5$ <$z $<$ 0.7 galaxies (star symbols) from sample B; the solid line 
was drawn from
the study of local galaxies by K92, the dotted lines delimit the area
containing all the local normal galaxies.  Error bars are only shown
for objects with $W_{0}(OII)>$0 for cosmetic reasons.}

\end{figure}

K92 discussed the relation between the ``red" index D(41-50) and
$W_{0}(OII)$, finding a very well-defined sequence for nearby
galaxies.  In passing, he mentioned that objects much redder in
D(41-50) than the defined sequence are likely to be Seyfert2 galaxies
(see Section 4.2).  Figure 11 shows the K92 diagram for our CFRS
galaxies with z $<$ 0.7.  The K92 sequence is shown by the solid line,
and the two dashed lines basically delimit the area containing all
K92's normal galaxies. Figure 11 demonstrates that there are objects at
all redshifts significantly redder than the K92 sequence and that at
higher redshift (z $>$ 0.5), there are galaxies (star symbols) having
 D(41-50) significantly bluer than the K92 sequence, while K92 found no 
such objects locally. This is related to the fact that the decrease of
D(41-50) index with redshift is accompanied by a significant increase
of the dispersion (Figure 7). In this section, we test if the apparent
evolution of D(41-50) color index versus $W_{0}(OII)$ is accompanied by
a change in the properties of the emission-line galaxies. Between z =
0.375 and z = 0.625 the luminosity function of blue CFRS galaxies is
found to be evolving by $\sim$ 1 mag around $M_{B}= -20.5$, if the
evolution is assumed to be in luminosity (CFRS VI).

For z $<$ 0.7, the most prominent emission lines found in our spectral
window are [OII] 3727, H$\beta$ and [OIII] 5007.  Photoionization
models (Stasinska 1984; Osterbrock 1989) show that the corresponding
line ratios depend on the nature of the ionizing continuum, the stellar
temperature, the metallicity and extinction. Unfortunately, our spectral
resolution does not allow accurate correction of the stellar absorption
underlying the H$\beta$ line. Therefore, we have proceeded as
follows:\\
(1) the underlying H$\beta$ absorption has been estimated from BC93
models using the relationship between W(H$\beta$) and the D(3550-4150)
index, two quantities which are known to be rather insensitive to the
metallicity (Bica \& Alloin 1986, Bica et al. 1994). 
 It has been estimated from the averaging of three different
models (1Gyr burst, 0.1Gyr burst of 10\% and of 50\% strength
percentage occuring in a 10Gyr old galaxy), and can be approximated by two
line segments with an accuracy of $\sim$ 1\AA\ on W(H$\beta$). 
From this relation we calculate the underlying 
H$\beta$ absorption and then correct the
H$\beta$ emission line. After correction, we found all W(H$\beta$) but
three were positive (i.e.,  in emission). For the three remaining
spectra, we have assumed a positive value for $W(H\beta)$, which is the
maximum of (1\AA, errorbar).\\
(2) since most of the objects have emission lines, it is possible
 to use the diagnostic diagram  for all objects having both [OII] 3727 and
[OIII] 5007, i.e., 60\% of the z $<$ 0.65 objects. The emission
line ratio $[OIII]_{4959+5007}/H\beta$ has been estimated by the ratio of
$W_{0}(OIII)$ to the corrected $W_{0}(H\beta$), while for the ratio
$[OII]_{3727}/H\beta$, the approximation used is:\\
\noindent
$[OII]_{3727}/H\beta = \\
 ([OII]_{3727}/W_{0}(H\beta) * (W_{0}(OIII)/[OIII]_{5007})$.\\
This relation simply assumes that the continuum at $H\beta$ can be
approximated by the continuum at 5007\AA.

\subsection{Diagnostic diagram and the fraction of AGNs}

It is very useful to try to identify AGNs since, besides being
interesting in their own right, any study of star formation activity is
likely contaminated by such objects, which are driven by other
mechanisms than ionization by stellar continua. We have used the
diagnostic diagram of ($[OII]_{3727}/H\beta$ vs $[OIII]_{4959+5007}/H\beta$)
 described in
Rola (1995) or Tresse et al. (1996, hereafter CFRS XII), and compared
the results to photoionization models (Rola 1995).
\begin{figure}[tbp] \label{f12}
\plotone{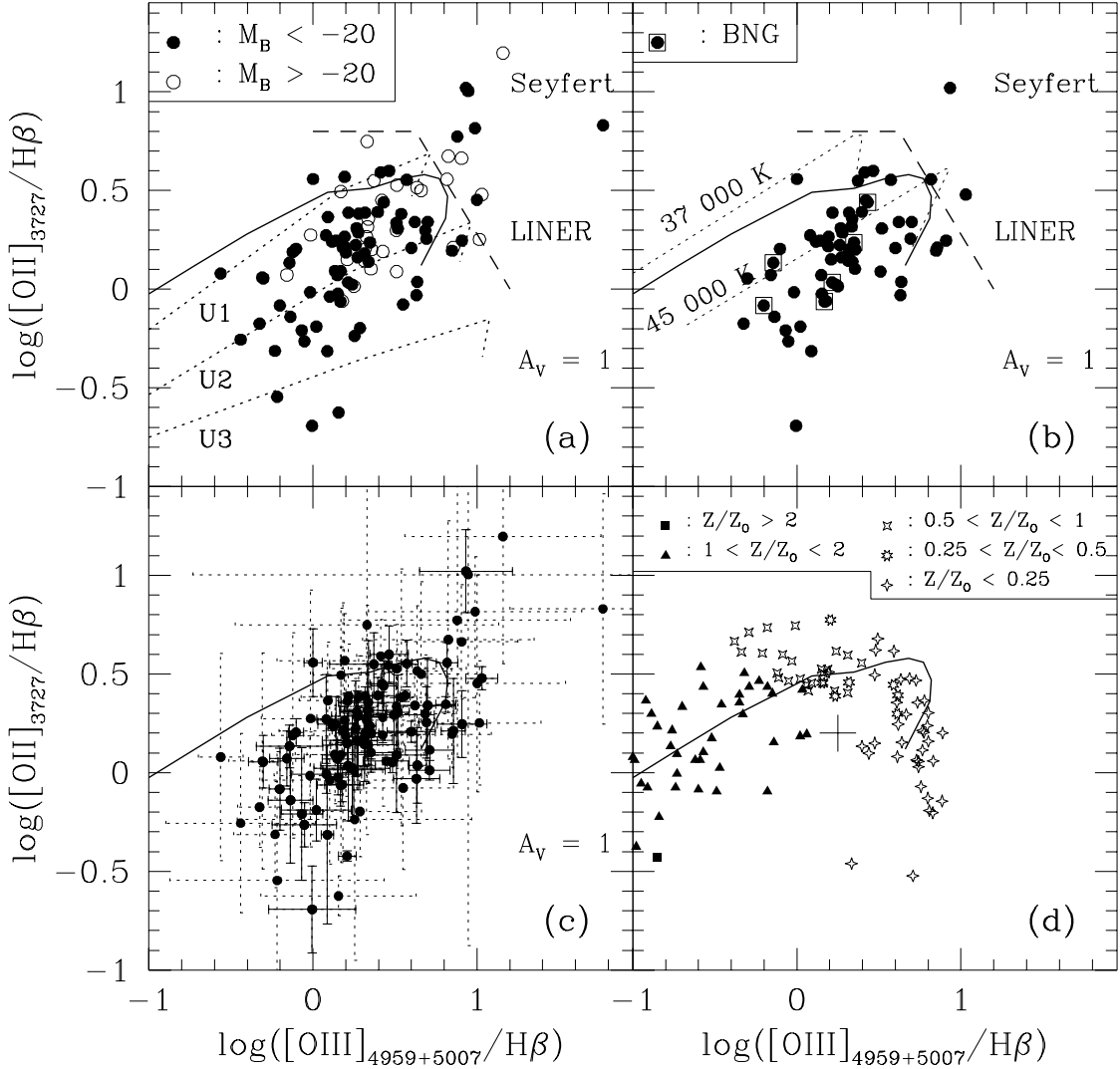}
\caption[]{Diagnostic diagram with $[OIII]_{4959+5007}/H\beta$
versus $[OII]_{3727}/H\beta$. {\it (a)} For 102 emission line galaxies
 (sample I) with z $<$0.7 (full dots: $M_{B} < -20$, open dots: $M_{B} 
> -20$). An average
extinction of $A_{V}$= 1 has been assumed. The H$\beta$ intensities
have been corrected for underlying absorption (see text). The solid line 
shows the theoretical sequence from McCall et al. (1985)
which fits local HII galaxies well, with metallicity decreasing 
from the left to the right. The dotted lines are drawn for three 
ranges of the ionisation parameter: U1, from 1.2 $\times$ 10$^{-3}$ to
4.7 $\times$ 10$^{-4}$ and Q$_{H}$ $=$ 1.2 $\times$ 10$^{50}$ s$^{-1}$; U2, from 7.4 $\times$ 10$^{-3}$ to
2.5 $\times$ 10$^{-3}$ and Q$_{H}$ $=$ 1.2 $\times$ 10$^{50}$ s$^{-1}$; U3, from 3.9 $\times$ 10$^{-2}$ to
1.2 $\times$ 10$^{-2}$ and Q$_{H}$ $=$ 1.2 $\times$ 10$^{52}$ s$^{-1}$. 
The dashed line shows the photoionization limit for a stellar 
temperature of 60 000K and empirically delimits the Seyfert 2 area 
from the HII region area. {\it (b)} For 70 galaxies with error bars
smaller than 0.3 in logarithmic scale. The dotted lines are drawn 
with varying metallicities for two given temperatures (from Mc Call 
et al. 1985). Blue nucleated galaxies are also displayed. {\it (c)}
Same as (a) with the error bars. {\it (d)} For a local sample of
HII galaxies with varying abundances (see references in the text). The
average value from the CFRS HII galaxy combined spectrum is also
 displayed (cross) for comparison.}
\end{figure}

Figure 12 shows the diagnostic diagram for all the 102 galaxies with
[OII] 3727 and [OIII] 5007 in emission (among 162 having defined 
D(3550-4150) and
emission line indices). The dashed line (panels (a) and (b)) shows 
an empirical limit
between AGN-like and stellar-like spectra, and corresponds to the
photoionization limit for a stellar temperature of 60 000K (see Rola
1995; CFRS XII). We have assumed an average extinction of $A_{V}$$=$1
mag (average value for HII regions, see Oey \& Kennicutt 1993), but
find that the results are not strongly dependent on that. An extinction
of 1 mag corresponds to 0.33 mag or 0.13 in $log([OII]_{3727}/H\beta)$, less
than the error bars. The result is that 11$\pm$6 objects among 162 z
$<$ 0.7 galaxies have Seyfert2-like spectra (we find no low excitation
AGNs, i.e. LINERs). Figure 13 shows their combined spectra with
prominent [OII] 3727 and [OIII] 4959 and 5007 emission lines. CFRS14.1567 
(z = 0.47, $M_{B}= -22$), which is likely a Seyfert 1 galaxy (Schade et al.
1996a), must also be added to the Seyfert sample at z$<$ 0.7.  The
fraction of Seyfert-like spectra is thus (12$\pm$ 6)/162, i.e. 7.4
($\pm$3.6)\% of the spectra.  This fraction is higher for the $M_{B} <
-20$ galaxies and might increase with redshift (11\% at z $>$ 0.6).
This result concerns objects which are brighter and at higher
redshifts than those discussed in CFRS XII, which describes properties of low
luminosity galaxies (z$<$ 0.3 and $-19.5<M_{B}< -17.5$), and found that
a significant fraction of the latter have emission-line ratio 
rather similar to those of LINERs.

\begin{figure}[tbp] \label{f5}
\plotone{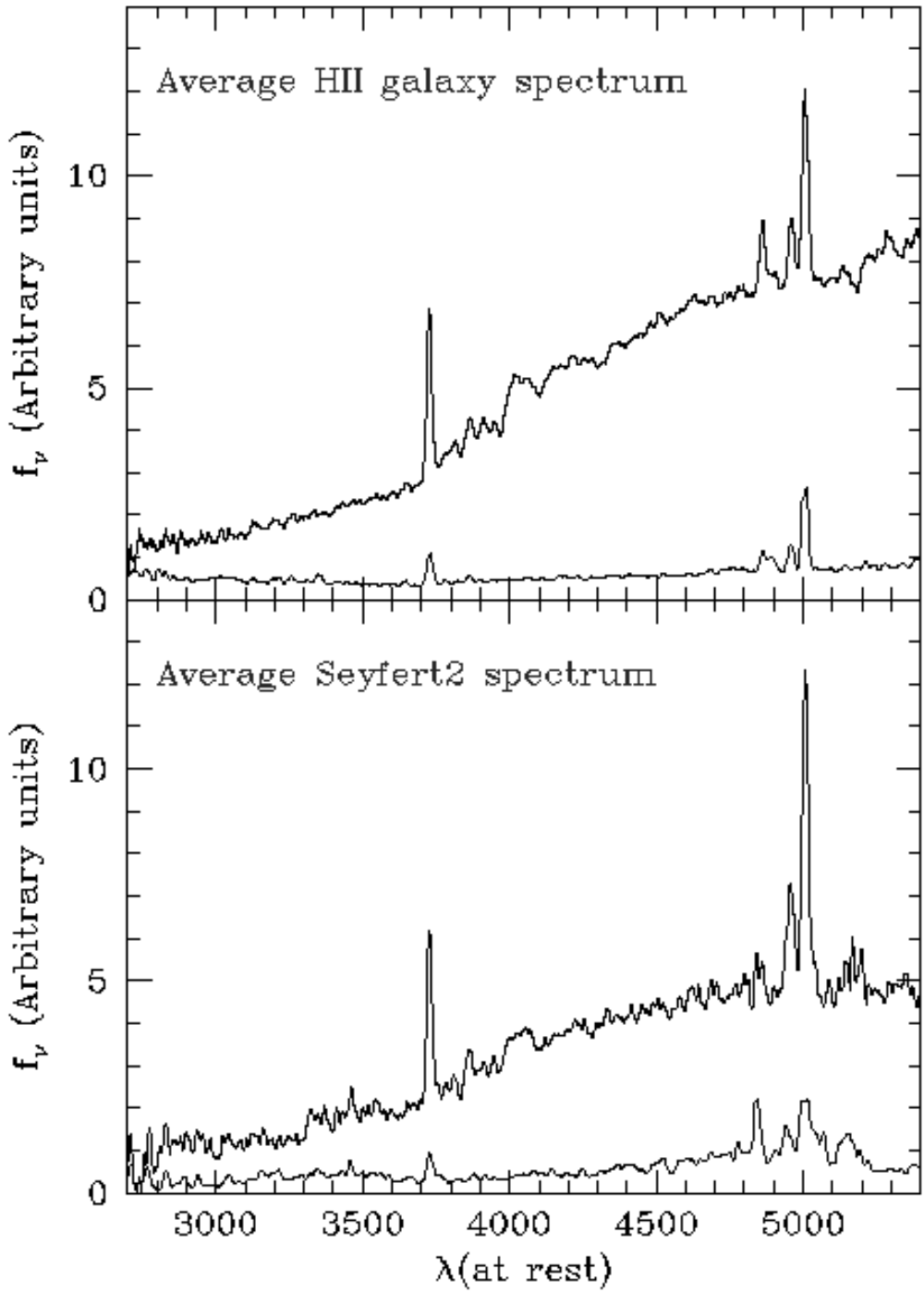}
\caption[]{{\it (top)}: Average spectrum of the 90 z$<$ 0.7 CFRS HII
galaxies. It shows a relatively red galaxy with substantial Balmer
absorption ($W_{0}(H\delta)$= 2.7\AA) and emission line ratios
(log($[OIII]_{4959+5007}/H\beta$)= 0.25 and log($[OII]_{3727}/H\beta$)=
0.22, assuming a H$\beta$ underlying absorption of 2.7\AA\ and
$A_{V}$= 1 mag), significantly discrepant from local HII region values.
{\it (bottom)}: Average spectrum of the 12 Seyfert2-like spectra which shows
the extremely strong $[OIII] 5007$ line relative to H$\beta$. The
latter is also probably affected by underlying absorption
($W_{0}(H\delta)$= 2.2\AA). The galaxy continuum is relatively red,
similar to several local Seyfert2 galaxies. Residuals 
(1$\sigma$ deviation) are shown in the bottom of each panel.}

\end{figure}

 The comparison of the fraction of Seyfert galaxies found in the CFRS
to that found in local samples is rather problematic, since the methodology
to identify Seyfert2 galaxies depends on the available emission lines in
the spectral window. For example Salzer et al (1989) find that 11\% of 
the emission line galaxies are Seyfert 
galaxies, on the basis of ($[NII] 6583/H\alpha$, $[OIII] 5007/H\beta$) 
diagnostic diagram. The other problem is the absence of systematic study
of the fraction of emission line in complete sample of low redshift
galaxies. However, there are some indications that the fraction of Seyfert
in local samples is close to 2\%, a value found by Huchra \& Burg (1992)
and which can be also reached by assuming that the rate of emission line 
galaxies locally is close to 20\% (from Schectman et al, 1992 or from 
Peterson et al, 1986) , and that 11\% of them are Seyfert as derived by
Salzer et al (1989). It indicates that the the fraction of CFRS galaxies 
that are Seyferts is probably much higher than that is found at low redshift.   
However, the fraction of emission-line galaxies that are
Seyferts appears to have stayed roughly constant at $\sim10\%$.

\subsection{[OII] 3727 versus continuum and AGN}

K92 mentioned that the $W_{0}(OII)$ vs D(41-50) diagram allows good
separation of AGNs and normal galaxies. He argued that ``the Seyfert2
nuclei tend to occur in early-to-intermediate type spirals which have
much redder colors than star-forming galaxies".  Among the 11 CFRS
galaxies with Seyfert2-like spectra, only one has a D(41-50) index
significantly redder than the K92 sequence line. One possibility is
that the mean continuum index becomes bluer with increasing redshift,
and hence the high-z Seyfert2 galaxies tend to lie closer to the
K92 line.

We have looked at objects having [OII] 3727 in emission and significantly
redder than the fiducial relation between $W_{0}(OII)$ and D(41-50)
color index ( $>$ 1 sigma from the dotted lines in Figure 11).  To be
more conservative, we added the condition that the objects must also
be significantly redder than the fiducial relation between $W_{0}(OII)$
and D(3550-4150), meaning that the whole spectral distribution should
be red from 3550 to 5000\AA, while having significant [OII] 3727 emission.
We identify six such objects (among 190 objects having both spectral
indices defined), and they generally show low ionization spectra
suggesting that they are LINERs. However, these six objects cannot be
used to estimate the fraction of LINERs at relatively high z, since
such objects are not easily detected at high redshift (moderate
$W_{0}(OII)$, no [OIII] 5007 lines).

\subsection{The nature of HII regions in galaxies to z = 0.7}

60\% of the z $<$ 0.7 galaxies show both [OII] 3727 and [OIII] 5007 emission
lines. Figure 12 shows their location in the ($[OII]_{3727}/H\beta$,
$[OIII]_{4959+5007}/H\beta$) diagram. We have adopted a uniform
extinction law, $A_{V}$= 1 (average value for local HII regions, see
Oey \& Kennicutt 1993). The solid line shows the model sequence of
McCall et al. (1985) which fits nearby extragalactic HII regions
very well. The sequence corresponds to varying temperatures and metallicities
while the dotted lines in panel (b) (also from McCall et al.) show the 
effect of
varying the metallicity (higher metallicities to the left) for a given
stellar temperature. Most of the CFRS emission-line galaxies  lie in a
region where no local HII regions are found. It would require 
extremely high extinction ($A_{V} >$ 3 on average) to bring the average
point of CFRS HII regions to the McCall et al. sequence. For a given value of
$[OIII]_{4959+5007}/H\beta$, the CFRS galaxies exhibit lower values of
$[OII]_{3727}/H\beta$ than local HII regions. Note also that luminous 
emission-line galaxies ($M_{B} < -20$) occupy the whole range of the diagram,
while less luminous galaxies (open circles) 
are absent in the leftmost part of the diagram.
This might indicate that HII regions with higher abundances are found in
more luminous galaxies, in agreement with the suggestion by Salzer
et al. (1989) from a study of nearby emission-line galaxies.

We have compared the emission line properties of the z $<$0.7 CFRS
galaxies to those of a local sample of HII regions and HII galaxies,
where the emission line intensities and corresponding oxygen abundances
were taken from the literature (Vilchez et al. 1988; Diaz et al. 1991;
Zaritzky et al. 1994; Pagel et al. 1992; Dinerstein \& Shields,
1986; Thuan et al. 1995; Skillman et al. 1989; Oey \& Kennicutt 1993;
Masegosa et al. 1994).  
 Only few CFRS HII galaxies have locations which 
coincide with that of local HII regions with abundances varying
between 1.5 and 0.05(O$/$H)/(O$/$H)$_{\odot}$ (or less).

We have computed photoionization models from the grid for HII regions
of Rola (1995; see also CFRSXII). These models were calculated using
Kurucz (1992) model atmospheres for a stellar effective temperature of
50 000K. The metallicity ranges from 2.0 to 0.1 times solar
metallicity.  The ionisation parameter is $ U = Q_{H} /(4 \pi R^{2}
n_{H} c)$, where $Q_{H}$ is the number of $H^{o}$ ionising photons -
which is 1.20 $\times$ 10$^{50}$ s$^{-1}$ in the two upper curves (U1
and U2) and 1.20 $\times$ 10$^{52}$ s$^{-1}$ in the bottom one (U3),
against 1.0 $\times$ 10$^{51}$ s$^{-1}$ for the McCall models -, R is
the radius of the photoionised region and c is the speed of light; U
increases from the upper curve to the bottom one (see Figure 12
caption). For each curve U1, U2 and U3, the ionisation parameter
slightly varies along each one, decreasing from the left to the right.
The lowest values of the ionization parameter reproduce quite closely
the observed local HII region sequence, whereas the highest possible
ionization parameters are required to match a significant fraction of
the CFRS HII galaxies (Figure 12a). In the models
which fit the CFRS objects, the ionization parameter ranges from
$\sim$5.0 $\times$ 10$^{-3}$ to 3.0 $\times$ 10$^{-3}$ approximately,
while the metallicity varies roughly between about 1.5 and 0.5 solar.
However the degeneracy of the models in this diagnostic diagram
prevents us from making firm conclusions about metallicity.

The upper panel of figure 13 presents the average spectrum of the 90
z $<$ 0.7 emission-line galaxies which are not Seyfert2. On average, the
[OIII] 5007 line has a higher flux than the [OII] 3727 line, conversely 
to what
is observed for local HII regions. Assuming that the underlying stellar
absorption is given by the $H\delta$ line (2.7 \AA), the combined
spectrum has an emission line ratio well off the McCall et al.
sequence (Figure 12d). This supports the fact that distant HII regions have
different physical properties from local ones.

In summary, the overall range in metallicities in CFRS HII regions in
galaxies at z $<$ 0.7 seems to be rather normal, although a substantial
fraction of these objects appear to have metallicities significantly
lower than the Sun.  The main difference between the local HII region
sequence and the CFRS sequence appears to be caused by a higher
ionization parameter for the latter. The effect seems to be stronger
with increasing redshift.  Many explanations can be proposed for the
redshift evolution of the ionizing properties in HII galaxies. It could
be due to an increase of the effective temperature at higher $z$, or to
a better ``efficiency" of ionizing photons coming from hot stars, the
latter possibly being related to a decreasing internal opacity between 
hot stars
and gas in HII regions of high redshift galaxies.
 
\subsection{H$\alpha$ versus [OII] 3727}
\begin{figure}[tbp] \label{f14}
\plotone{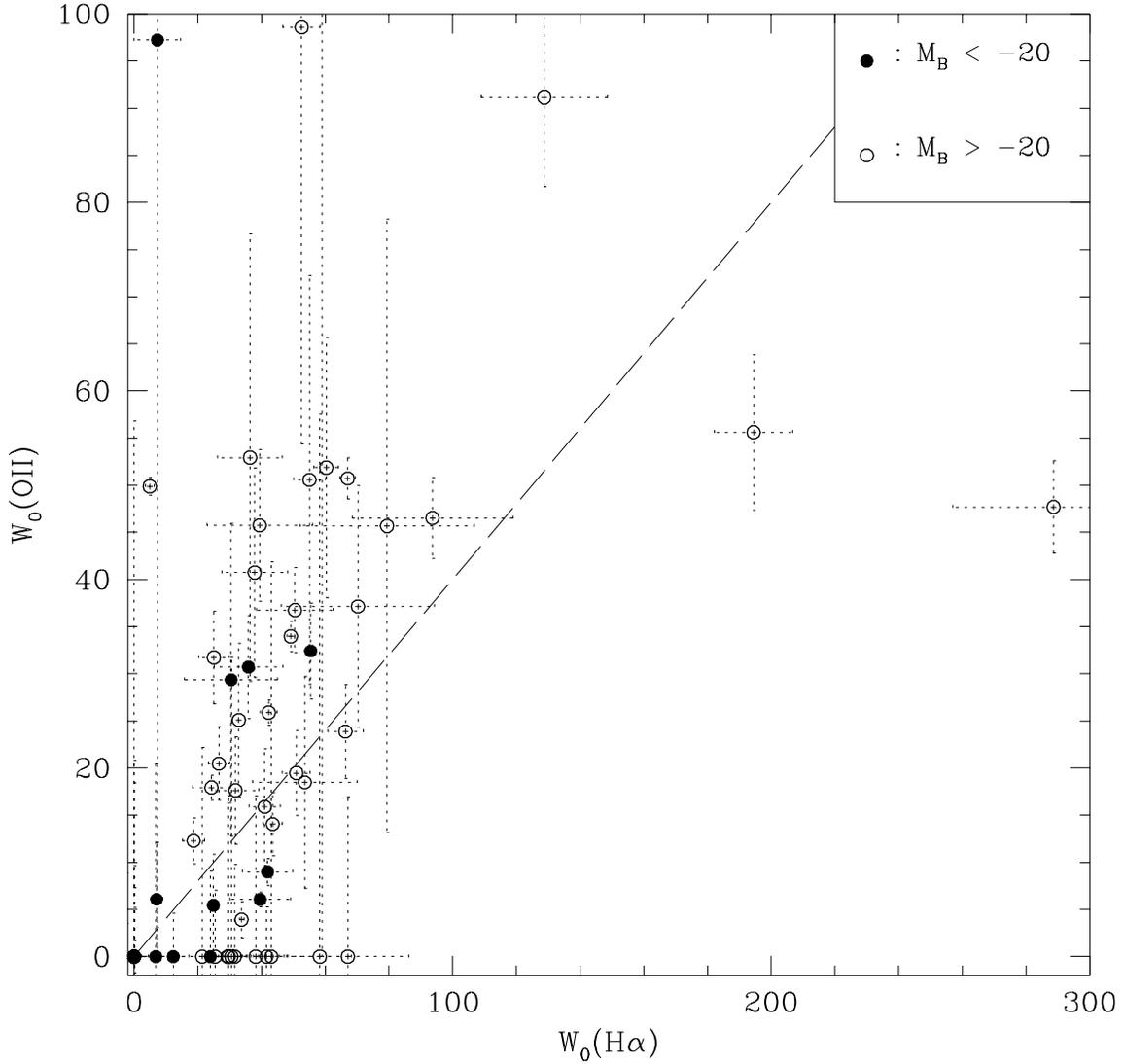}
\caption[]{ Relationship between [OII] 3727 and $H\alpha$ equivalent widths
for 61 CFRS galaxies (sample B) with z$<$ 0.3 (full dots: $M_{B}< -$20, 
open dots:$M_{B}> -20$). The dashed line shows the same relationship for the
K92 local galaxies. The lines are derived from the two opposite ends of the
spectrum which could lead to systematic errors in the flux calibration,
although equivalent width measurements are found to be largely
independent of the flux calibration.}

\end{figure}

Kennicutt(1992) reported a tight relationship between $W_{0}(OII)$ and
W($H\alpha$) in a local sample of galaxies and used it for calibrating
the star formation rate against the [OII] 3727 luminosity. A similar figure
(Figure 14) for the 61 CFRS galaxies at $z < 0.3$ shows a large
dispersion and  also shows that a large fraction of the objects with
strong $W_{0}(OII)$ lie well above the K92 local relation.
Several reasons might be responsible for this difference:\\
(i)  the CFRS galaxies at z$<$ 0.3 have luminosities 
($-20.5 < M_{B}< -17.5$) smaller on average than the population of local 
galaxies studied by K92 ( $-23 < M_{B} < -17.5$), and might present
less extinction (K92 used $A_{V}$= 1.26).\\
(ii) they might have metallicities somewhat lower than solar ($Z \sim
0.25 Z_{\odot}$), which can give larger $[OII]_{3727}/H\beta$ ratios (and
consequently the $[OII]/H\alpha$ ratios); for example,
Z$/$Z$_{\odot}=$0.25 would increase $[OII]_{3727}/H\beta$ by $\sim$ 1.8
(compared to Z$/$Z$_{\odot}=$1), assuming hot star temperatures (T $>$
40 000K, see Rola 1995).\\
(iii) a large fraction of our z $<$ 0.3 galaxies are low luminosity
objects with AGN activity (LINERs, see CFRS XII).\\
In view of these results, it is not obvious that K92's
relationship for nearby galaxies can be extrapolated to higher redshift
and, in particular, whether $W_{0}(OII)$ can be used to accurately
estimate star formation rates.  This will be discussed in more detail
in section 9.

\section{CONTINUUM PROPERTIES OF GALAXIES TO z = 1}

We have made extensive use of the BC93 code with
various IMF and star formation scenarios (single burst, burst of
constant duration etc.,). The deficiencies of the code have been
recently discussed by Charlot (1995), who compared it with other codes
and found discrepancies as high as 0.05 mag in ($B-V$) colors and 0.25
mag in ($V-K$) color. For all important issues related to the
modelling, we have compared our results with both the BC93 and BC95 codes.
The main concern with these models is that they are based on stellar tracks
with solar metallicities for all stellar ages.
\begin{figure}[tbp] \label{f5}
\plotone{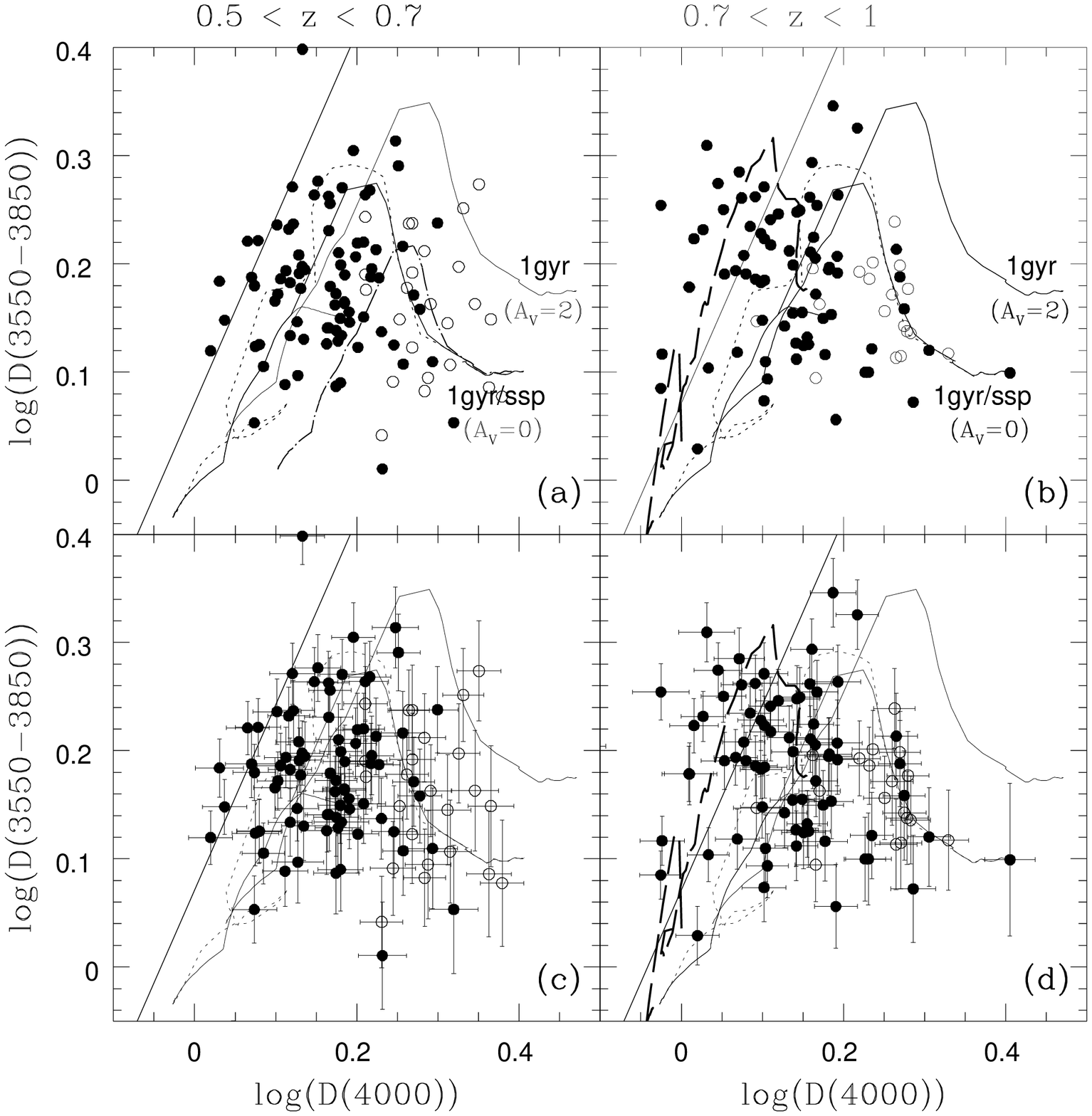}
\caption[]{ Balmer index versus D(4000) index (sample E). {\it(a)}: for
$M_{B}< -20$ with 0.5 $< z <$0.7 (full dots: $W_{0}(OII)>$ 0\AA; open 
dots: $W_{0}(OII)=$ 0). The lines show the tracks (age increasing from the left
to the right) from the population synthesis model of BC93, assuming a 
single burst (dotted line), or a burst of 1 Gyr duration (solid line).
A simple model of a 10\% strength burst occuring in a 15 Gyr galaxy is
also shown for illustration (long dash-dotted line).  
A model with extinction for a burst of 1 Gyr duration is shown to 
illustrate the current extinction vector (solid line for $A_{V}$= 2).
Basically all the points can be fitted by the BC93 models, or by 
a combination of them, assuming extinction $A_{V}\le$ 2).  
Very few points are found on the
left side of the dotted line which empirically delimits the D(4000)-deficient
galaxies. {\it (b)} Same as (a), for the 95 high redshift galaxies 
($M_{B}< -20$
with 0.7 $< z <$ 1). Conversely to (a), a significant fraction of
galaxies are found on the left side of the dotted line ( D(4000)
deficient galaxies), and cannot be fitted by any combination of the
BC93 or BC95 models with or without extinction.  The dashed line shows
the tracks for an instantaneous burst for a metallicity of 1/50 of the
solar value (from Bruzual and Charlot, 1996, dashed line). Models with
lower metallicity than solar provide an excellent fit to the
D(4000)-deficient galaxy.
{\it (c)} Same as (a) with error bars. {\it (d)} Same as (b) with error bars. }

\end{figure}
\begin{figure}[tbp] \label{f15}
\plotone{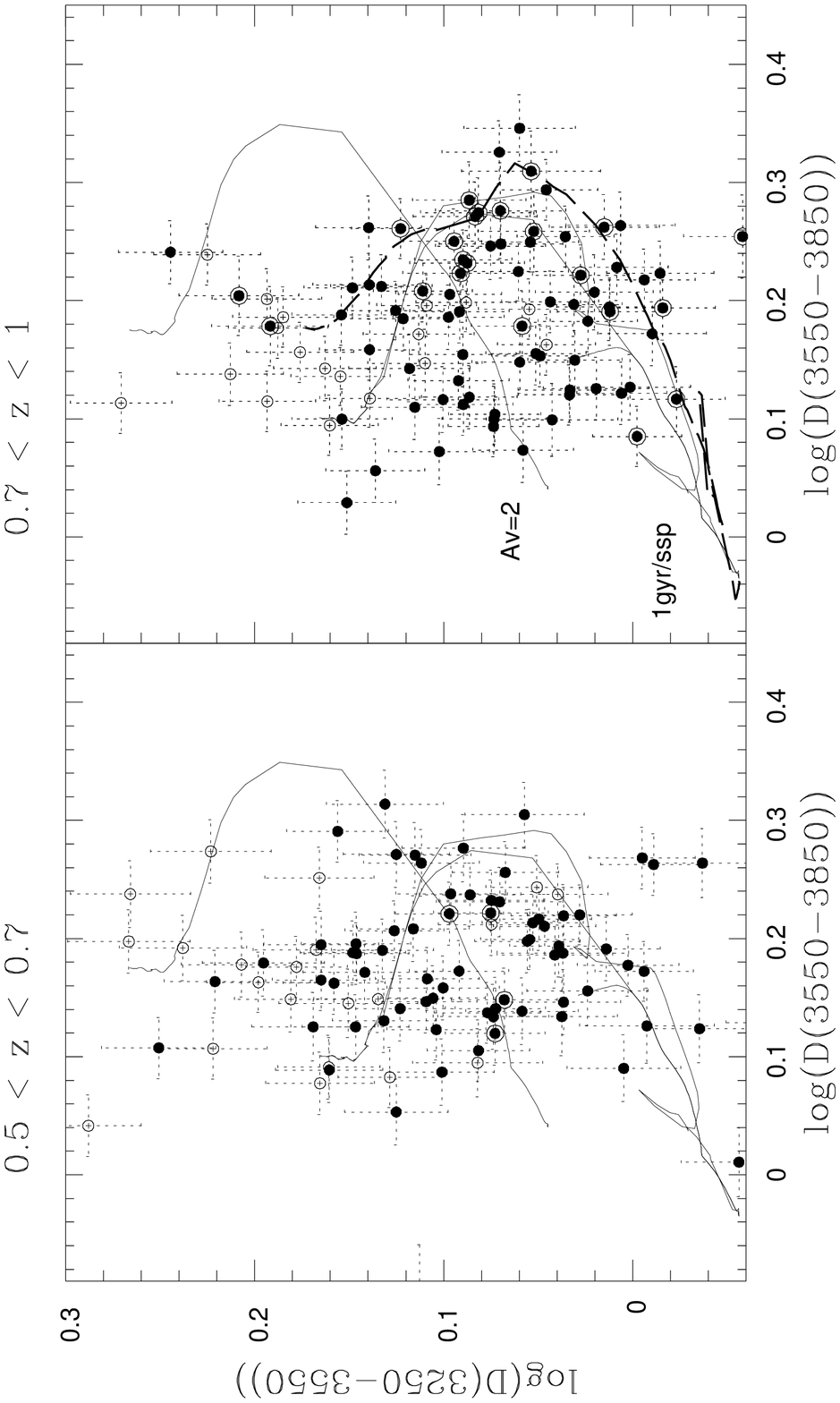}
\caption[]{ Balmer index D(3550-3850) versus UV D(3250-3550) index (sample E).
{\it (a, left)} for $M_{B}< -20$ galaxies with 0.5 $< z <$0.7 (full 
dots: $W_{0}(OII)>$ 0\AA; 
open dots: $W_{0}(OII)=$ 0\AA; circled dots: D(4000)-deficient galaxies). 
 Typical error bars are $\pm$0.03 in logarithmic scale. As in
Figure 15, solid lines show the tracks from the population synthesis
model of BC93 (increasing UV index from young to old stellar
population), with various extinctions ($A_{V}$=0 or 2). Almost all the
points can be fitted well by BC95 models, if one assumes various star
formation scenarios and extinctions. Notice the presence of emission-line 
galaxies which are redder than the oldest stellar population, and
which could be either strongly-absorbed young burst or a ``mini-burst"
occurring in an already-old stellar population (see section 8.3).
{\it (b, right)} Same as (a), for the high redshift galaxies ( $M_{B}< -20$ 
with 0.7$< z <$ 1). In contrast to what is found in the the Balmer versus
D(4000) diagram (Figure 15b), all the galaxies have their UV properties
well fitted by BC93 models (including with moderate extinctions). The dashed line 
shows the track for an instantaneous burst for a metallicity of 1/50
of the solar value (from Bruzual and Charlot, 1996). No
significant difference are found between models with various
metallicities, which explains why D(4000)-deficient
galaxies have their UV properties  fit well by solar metallicity
models. }

\end{figure}

In the following diagrams (Figures 15 and 16), we have classified the
CFRS galaxies in two main categories depending on the [OII] 3727
emission (open circles:  $W_{0}(OII)=$ 0\AA; full circles: $W_{0}(OII)
>$ 0\AA).  Furthermore,  diagrams are plotted for high redshift (0.7
$<z <$ 1.0) and intermediate redshift (0.5 $<z<$ 0.7 and $M_{B} < -20$)
subsamples.  This empirical division has several advantages: (i) there
is considerable evolution displayed by the luminosity function (CFRS
VI) between these two redshift bins (1 mag if assumed to be luminosity
evolution); (ii) the most luminous galaxies ($M_{1\mu} < -22$) show
large evolution of their colors and [OII] 3727 properties between these
two redshift bins (section 3.3); (iii) there are an equal number of
emission-line objects in the two redshift categories; (iv) it allows us
to investigate the evolution of important color indices such as D(4000)
and Balmer indices relative to emission line properties (section 3.3);
(v) in the low redshift bin, we have a fair representation of the
contamination by AGN-like spectra.

\subsection{The Balmer/D(4000) index diagram}

The D(4000) index is well known to be dependent on both temperature and
metallicities (see Worthey 1994).  The BC93 models (with solar
metallicity) show that both the D(4000) and Balmer D(3550-3850) indices
increase with age up to $\sim$0.5 Gyr, but then the Balmer index
decreases  due to the increasing fraction of old stars. The Balmer
index  is also somewhat dependent on metallicity, but it is mainly
age-dependent up to 0.4 Gyr (Bica, Alloin \& Schmitt 1994).  The
Balmer/D(4000) index diagram is not very sensitive to extinction for
the youngest ages, where the stellar tracks are nearly parallel to the
extinction vector. Figure 15a shows the z $<$ 0.7 galaxies ($M_{B} <
-20$) in such a diagram. Many of the galaxies are  fitted reasonably
well by the BC93 or BC95 models, especially if effects of extinction 
($A_{V}$ of 1 -- 2 mag) are included for a number of the quiescent objects.
There are several objects with emission lines which have intermediate
properties between old and young galaxies, including most of the
AGN-like objects.  They have either a lower Balmer index and/or a
higher D(4000) index than expected from the BC93 model. These
intermediate objects are likely the ``normal" galaxies in which star
formation takes place in a relatively old galaxy ($>$ 3 Gyr, see the
long dashed line in Figure 15a).

Figure 15b presents the same diagram for the highest redshift
galaxies.  There are fewer objects in the red, highly-absorbed part of
the diagram, but this might be related to selection bias against
quiescent galaxies at high redshifts (see section 7.1).  More
significantly, almost $\sim$ 1/3 of the emission-line objects have
extremely small D(4000) indices with relatively high values for the
Balmer index. These cannot be fitted by BC95 models even with extreme
IMFs or star-formation histories. Even assuming an IMF extremely
deficient in normal and low mass stars (i.e. a Salpeter IMF with lower
mass limit 2.5 $M_{\odot}$ and upper mass limit of 125 $M_{\odot}$),
only marginally helps the fit. We refer to these objects as
``D(4000)-deficient objects" and return to them in Section 6.

\subsection{The Balmer-UV diagram}

The D(3250-3550) index is mainly sensitive to very young populations
since metallic absorption lines are weak in this wavelength region.
Figure 16a shows the relation between the D(3250-3550) and D(3550-3850)
indices for the lower redshift bin.  A large fraction of the objects
are fitted by the BC93 models when extinction is included.  There are
several objects in the top left of the Figure 16a which deserve special
attention, since they simultaneously show  emission lines and a UV
color D(3250-3550) redder than that of an old stellar population (see
section 8.3). At z$>$ 0.7, Figure 16b shows that all objects can be
fitted by the BC93 models (with relatively modest amounts of
reddening), including the D(4000)-deficient objects.  
\begin{figure}[tbp] \label{f5}
\plotone{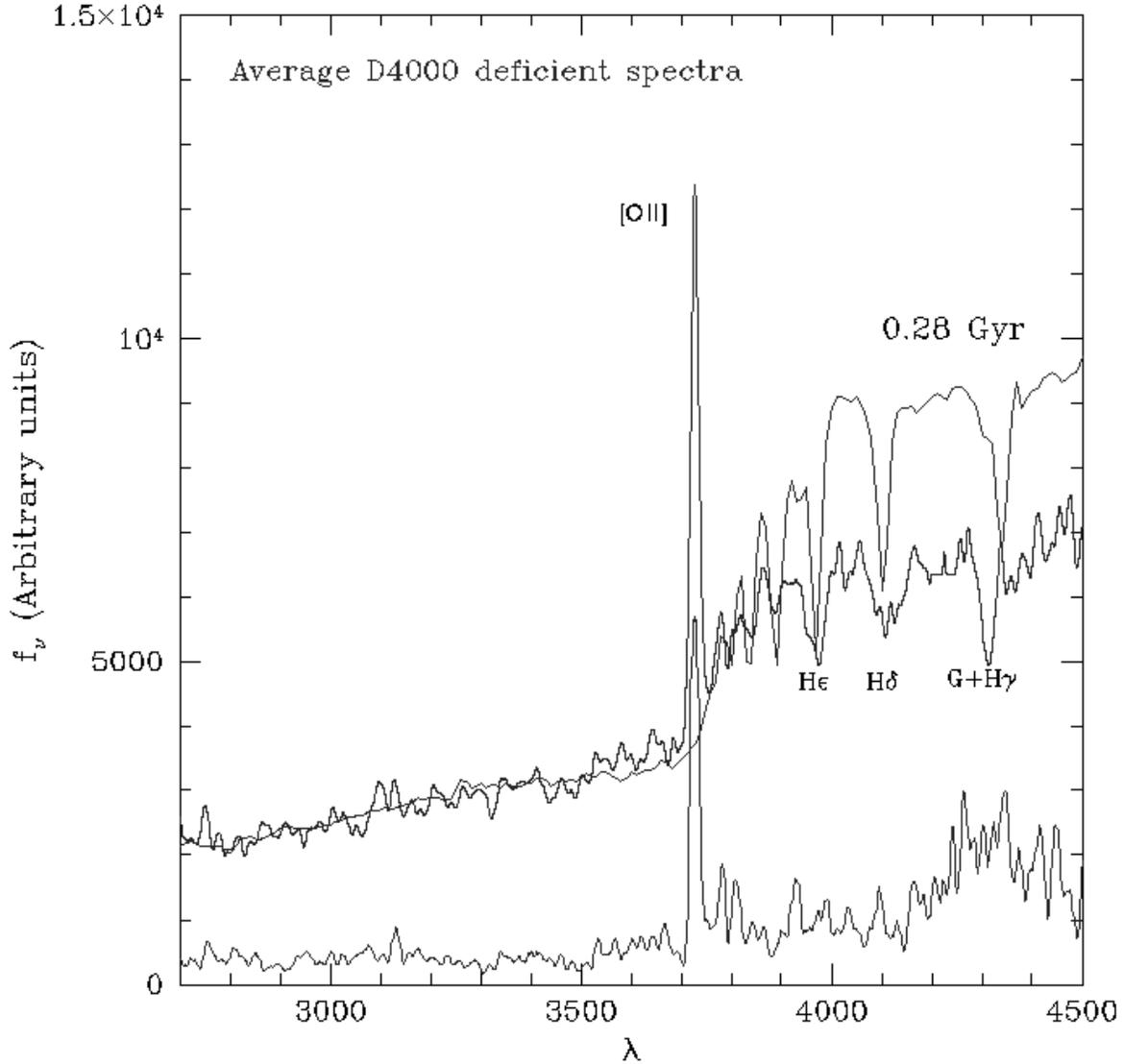}
\caption[]{ Thick line: average spectrum of the 23 D(4000)-deficient galaxies
at z$>$ 0.7. The large Balmer index, log(D(3550-3850) = 0.22 as well as
the Balmer absorption lines (equivalent widths of $H\delta$ and
$H\epsilon$ between 3 and 4\AA) indicate the presence of a
significant A star population. Thin lines: fit of the UV part of the average
D(4000)-deficient galaxy spectrum by a single burst template spectrum
from BC93 code. The latter fits the continuum  well below 3900\AA\ while it
overestimates the observed flux by a factor 1.5 above 4000\AA. Other
scenarios (constant burst duration, etc.,..) would not provide a better
fit of the whole spectral energy distribution of such objects from 2750
to 4500\AA. The age (0.28 Gyr) indicated by the single burst model, is a
low estimate since the D(4000)-deficient galaxies still show signs of
star formation. Residuals (1$\sigma$ deviation) are shown in the bottom.}

\end{figure}

\section{THE D(4000)-DEFICIENT OBJECTS}

Initially, we were concerned that the D(4000) deficiency noted in many
objects (Section 5.2 above) might be an
artifact in our data, but the reality of these features has been
confirmed by inspecting the individual spectra and by looking for
possible systematic effects that might have occurred at high
redshifts.  No such effects were identified and we believe their D(4000)
deficiencies are real.  Moreover the computations of D(4000) indices
are not significantly affected by sky emission lines (section 2.4) and
the D(4000)-deficient objects have no peculiar distribution in the
relation between spectral index and restframe colors derived from
spectroscopy (Figure 2). 
Support for the existence of this population
at high redshift is provided by published Keck
spectra (Cowie et al. 1995). 

The dashed line in Figure 15b empirically separates these objects from
the rest of the sample. The combined spectrum (Figure 17) of the 23
D(4000)-deficient objects shows relatively strong Balmer absorption
lines ($H\delta, H\epsilon$) indicative of the presence of an A star
population. The [OII] 3727 line is often very strong (on average,
$W_{0}(OII)$= 40\AA\ at rest), and a significant fraction of the
strongest [OII] 3727 emitters in the sample are D(4000)-deficient
objects (Figures 4, 9).

Almost all the D(4000)-deficient objects are fitted by the BC93 models
in the Balmer-UV diagram (Figure 16b) without requiring significant
extinction or bimodal stellar populations.  Their UV continua below
3650\AA\ is not as blue as expected for a very young population of
stars. If one fits the UV part (from 3100 to 3900\AA) of the spectrum
shown in Figure 17 by a single burst BC95 model, the light above
3900\AA\ is overestimated by a significant factor (from 0.3 to 0.5
mag). Age does not appear to be the dominant factor.
\begin{figure}[tbp] \label{f5}
\plotone{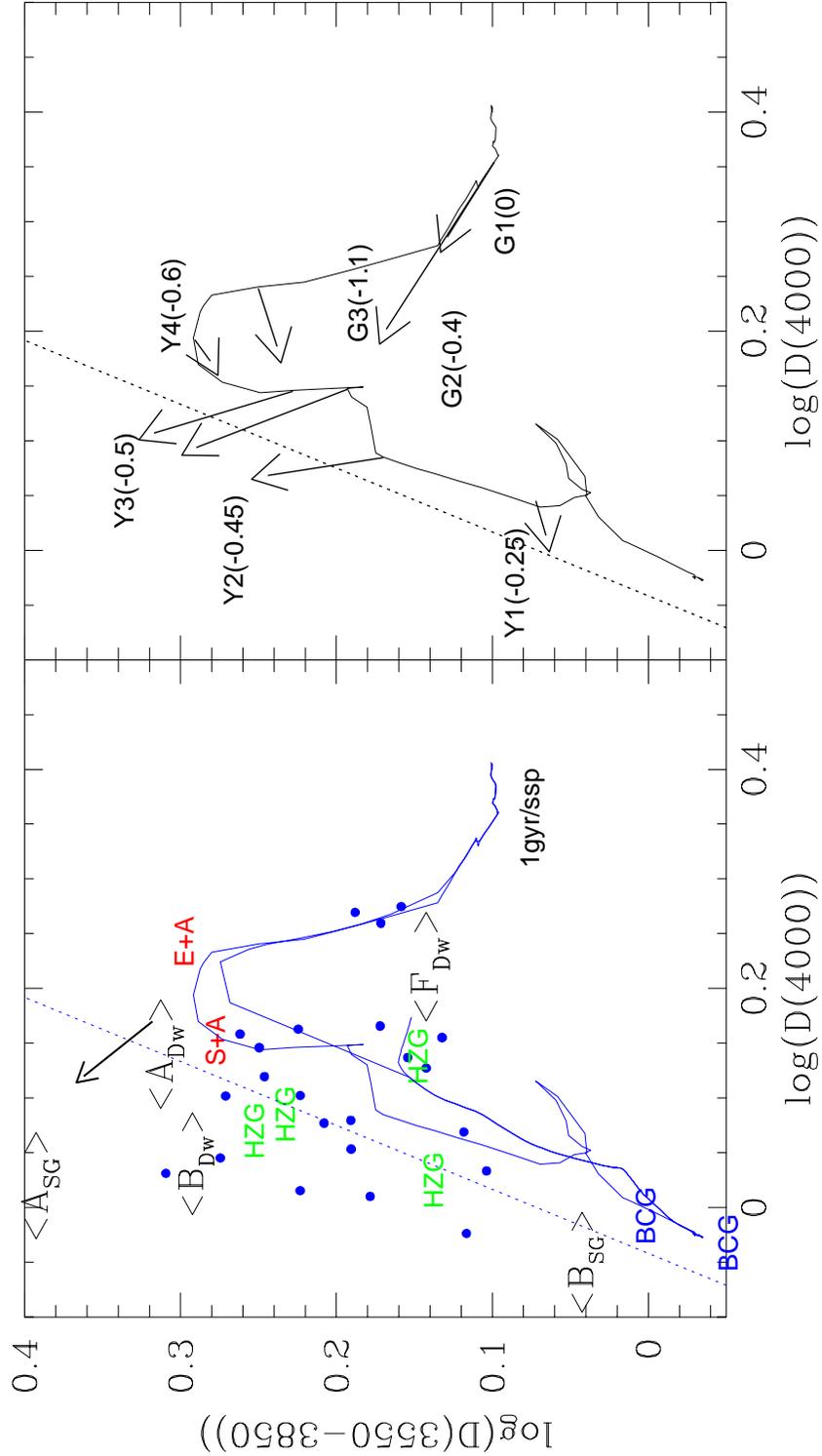}
\caption[]{ {\it (a, left)} Balmer versus D(4000) index as in
Figure 15, in which we have only included the 31 z$>$ 0.7 galaxies
with a very good agreement between their spectrophotometric and their
photometric colors (better than 0.1 magnitude, see section 2.3). The
location of A , B and F dwarf and supergiant stars are shown for
comparison. The arrow shows the effect of decreasing metallicity from
solar value to one tenth of the solar value, for an A star. The location of E+A,
S+A and of the blue compact galaxies is also shown. Added to the
fact that solar metallicity globular clusters are well fitted by the BC93
models, this figure demonstrates that population synthesis models
fit well the properties of young to old galaxies at low to high
redshift. HZG symbols show the location of the four galaxies from Cowie
et al. (1995) with a redshift ranging from z = 1 to 1.2. Three of them are
found in the same area as the D(4000)-deficient galaxies which suggests
that they share similar properties. {\it (b, right)} Balmer
versus D(4000) index  for young, intermediate and
globular clusters from Bica et al. (1994). We have assumed that star
clusters follow a single burst scenario, and relate the model
to the star cluster by an arrow. The effect of metal deficiency is
to decrease the D(4000) index while increasing the Balmer index.
Intermediate age and metallic deficient star clusters (from the
Magellanic clouds) are located on the right of the dotted line, a
location that they share with the D(4000)-deficient galaxies.}

\end{figure}

\subsection{Comparison to other astronomical objects}

Figure 18a shows the location of A, B and F
stars, from supergiants (higher Balmer indices) to dwarfs (from Silva
\& Cornell 1992), in the Balmer-D(4000) diagram. The presence of an
extra population of A or B stars in D(4000)-deficient galaxies might be
indicated  by their strong average $H\delta$ absorption (Figure 17).
In the near UV and blue, the D(4000)-deficient object light might come
from a mix of A and B stars without significant contribution of other
stellar populations. This would mean that the predictions from the
current population synthesis models are incorrect.  We do not believe
that this is the case because: (i) the BC93 models predict very well
the properties of D(4000)-deficient objects below 3900\AA; (ii) even A
supergiants have higher D(4000) indices than the  bluest half of the
D(4000)-deficient objects. Most of the Silva \& Cornell stars have
solar metallicity. However, the arrow shows how an A star would move in
the diagram when the metallicity decreases from solar to 0.08 times the
solar value. Among the four Cowie et al. (1995) spectra of galaxies with 1
$<z<$ 1.2, three lie in the D(4000)-deficient area of our diagram
(labelled as HZG in Figure 18a).

Figure 18b shows the comparison with the BC93 models and globular and
open star clusters from Bica et al. (1994). The Magellanic Cloud young
clusters (Y*) have ages from $10^{7}$ to 5 $\times 10^{8}$, the
globular clusters (G*) have ages from 7 to 15 $\times 10^{9}$ yr. The
young clusters have sub-solar metallicities while the  metallicities of
the globulars range from solar to sub-solar (the number in parenthesis
is the logarithm of Z$/$Z$_{\odot}$). We have assumed that star
clusters follow a single burst  scenario, and  demonstrate by an arrow
in Figure 18b how the Bica et al. clusters move from the BC93 model
predictions with solar metallicity. For globular clusters, one can
easily see that decreasing metallicity gives a higher Balmer index and
a lower D(4000) index. For younger clusters, all the vectors are in the
direction of decreasing the D(4000) index and generally of increasing
the Balmer index. Note that the young clusters (Y2 and Y3, single burst
ages from 0.05 to 0.1 Gyr) lie in the same area as D(4000)-deficient
objects. It should also be noted that the clusters have not been
corrected for intrinsic extinction. This would move them towards
smaller D(4000) and smaller Balmer indices.

Figures 18a and b demonstrate that other astronomical objects lie in
the same region of the D(4000)-Balmer diagram as the D(4000)-deficient
objects, including stars and intermediate/young  metal-deficient star
clusters.  The general effect of a lower metallicity on a
young/intermediate star population is to increase the Balmer index
while decreasing the D(4000) index. However, before reaching a definite
conclusion, one must investigate how BC93 models fit galaxies of
various types. We already know that below z = 0.7 almost all the CFRS
galaxies can be fitted by BC93 models with solar metallicity, sometimes
with additional assumptions such as extinction or bimodal burst
scenario (i.e. a second burst in an old galaxy). In Figure 18b one can
see that BC93 models fit globular clusters well with solar metallicity,
and hence, early type galaxies. Figure 18a shows a comparison between
high-z CFRS galaxies and other galaxies, such as the so-called E+A
(Dressler and Gunn, 1983), S+A (Hammer et al. 1995b) and the blue
compact galaxies (BCG, Izotov et al. 1995).  The agreement between the
BC93 models and the observations is remarkable, the differences being
much smaller than the discrepancy with the D(4000)-deficient objects.
Even the blue compact galaxies which have significant sub-solar
metallicities are fitted well. This is consistent with the fact that at
very young ages, the metallicity seems not to affect the color indices
of the star clusters (see cluster Y1 in Figure 18b).

\subsection{Comparison to models with low metallicity}

Metallicity has two main effects on the continuum properties. Firstly,
metals play a crucial role in radiative transfer (cooling) and
decreasing metallicity generally results in a blueing of the
continuum.  Secondly, metallic absorption lines can also affect the
continuum. Bica \& Alloin (1986), Bica et al. (1994) and Bonatto et al.
(1995) have done detailed studies of this from 1200 to 10000\AA\ using
star clusters. Population synthesis models with metallicities lower
than solar have also been recently developed (see Charlot 1996a for a
review).  The uncertainties between models from various authors have
been discussed for old stellar populations and young star-forming
galaxies by Charlot et al. (1996) and Charlot (1996a), respectively.
They find that predictions based on broad-band colors such as $B-V$ and
$V-K$ can be affected by substantial errors at  low ages (T $ < 10^{7}$
yrs), especially because of nebular emission. At intermediate and old
ages, the discrepancies among various models appear less severe.

The dashed lines in Figure 15b show the tracks of an instantaneous
burst with a metallicity $Z=0.02 Z_{\odot}$, from the very recent model
of Bruzual and Charlot (1996, kindly provided by Stephane Charlot). In the
Balmer/D(4000) diagram (Figure 15b), it is a relatively good fit to the
average properties of D(4000)-deficient objects.  In the UV/Balmer
diagram (Figure 16), the low metallicity model shows no significant
differences from the solar metallicity model.

Additional support for low metallicity of these galaxies comes from
Worthey's (1994) models.  Worthey (1994) estimated that if two stellar
populations differ in age and metallicity by d~log(age)/d~log(Z) $=$ 1.3,
they would present the same D(4000) index. The D(4000)-deficient
galaxies at high z have D(4000) = 1.04 $\pm$ 0.19, which would be
fitted by a BC93 model of 1Gyr burst with an extremely low age, 
7.25$\times 10^{6}$ years. The average UV index, which is nearly
independent of metallicity, is 1.15 $\pm$ 0.18, implying a much
older age, 1.2$\times$ $10^{9}$ years.  To bring the ages into agreement, a
metallicity of Z$/$Z$_{\odot}=$0.02 would be required, a value in good
agreement with that found from the BC models.

Since the low-metallicity models still require further tests, we adopt
a conservative upper limit for the metallicity of the D(4000) deficients
objects, $Z/Z_{\odot} <$ 0.2, which comes from
comparison with the  young clusters in the Magellanic Clouds.
\begin{figure}[tbp] \label{f5}
\plotone{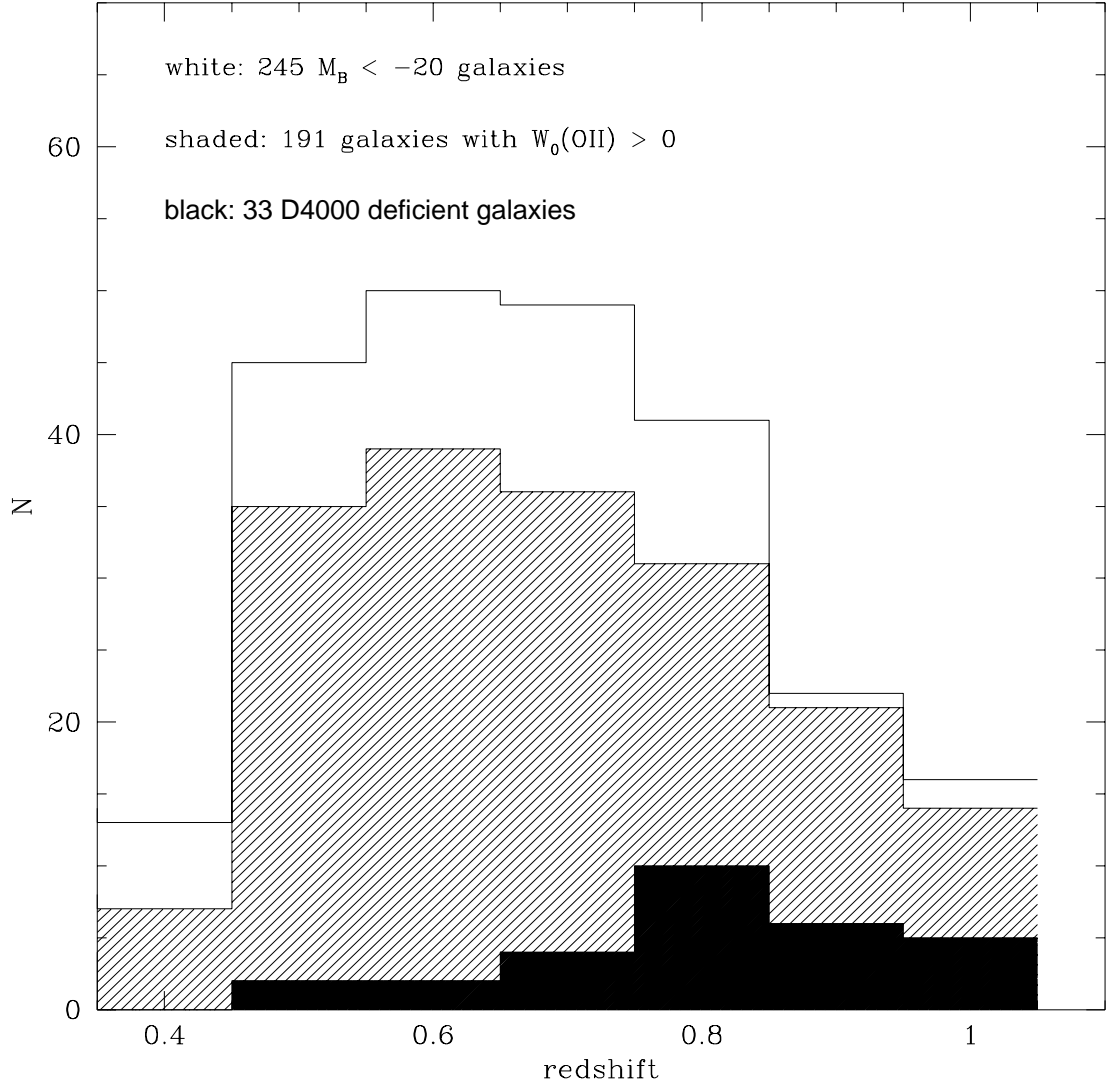}
\caption[]{ Histogram of the redshift distribution of 245 $M_{B}<
-20$ galaxies (sample E:0.35$< z <$ 1, white); 191 emission-line galaxies
(shaded); 33 D(4000)-deficient galaxies (black). The latter represent only a
small fraction of the galaxies at z$<$ 0.7, while they represent from
24\% to 31\% of the sample beyond z = 0.7.}
\end{figure}
\subsection{An emergent population of low metallicity objects at
high redshift?}

From the comparison with intermediate age star clusters and with low
metallicity models, there is considerable evidence that
D(4000)-deficient objects are actually metal-deficient objects. Figure
19 shows the distribution of these objects ($M_{B} < -20$) against z.
While they correspond to only a few percent of the sample at z $<$ 0.6,
D(4000)-deficient objects represent from $1/4$ to $1/3$ of the galaxy
population at z $>$ 0.7. 

These galaxies share several properties with most of the other galaxies
at high redshift, including their distribution in restframe luminosity
at 1$\mu$ (Figure 9). A large fraction of the z$>$0.7 galaxies with
very small $(U-V)_{AB}$ color indices and large [OII] 3727 luminosities are
D(4000)-deficient objects (Figure 5). The UV properties of D(4000)
deficient objects are well fitted by BC93 models, and most of them lie
very near the blue sequence of BC93 stellar tracks (ages lower than 1
Gyr), with no evidence for extinction. More than half of them have
log(D(3550-3850)) $>$ 0.2, implying a significant contribution of A
stars.  The low metallicities, if truly as low as indicated,
imply that the star formation took place
several $10^{8}$ years previously in an almost primordial medium.

We have examined available HST images (from the HST archive and from CFRS IX) 
for five D(4000)-deficient galaxies: CFRS03.0485, 14.0985,
14.1189, 14.1446 and 14.1496. Two show well-defined disks and two have
a very compact component (one of them is classified as ``blue
nucleated" in CFRS IX).

Low metallicities have been reported in most of the damped $Ly\alpha$ 
systems from z=0.7 to z=3.4 (Smith et al, 1996). Pei and Fall (1995) have 
shown that selection effects against systems with high dust obscuration 
are probably at work in samples of damped $Ly\alpha$ systems. Observed 
damped $Ly\alpha$ systems could be the most transparent ones, which 
raises many interests in comparing their properties to those of the 
D(4000) deficient objects.

\section{THE QUIESCENT GALAXIES}

\subsection{Evolution with the redshift}

The fraction of luminous quiescent galaxies ($M_{B}< -$20 and $W_{0}(OII)=$0) decreases
with redshift from 53\% at z = 0.3 to 23\% at z$>$0.5 (Figure 3).
Below z = 0.3, luminous quiescent galaxies likely include E/S0 and a
sizeable fraction of later type galaxies (Sa, Sb, etc.,.), assuming that
there is no large difference in the mix of galaxy population relative
to the multi-type luminosity function of Bingelli et al.  (1988).
Beyond z = 0.5, the fraction of quiescent galaxies reaches 23\% (Figure
3), and is in good agreement with the number of bulge-dominated
galaxies (7 E/S0) found in the small subsample of 32 galaxies with
z$>$0.5 observed with HST (CFRS IX). A careful analysis of the
spectra of the CFRS IX galaxies reveals seven quiescent galaxies,
including five bulge-dominated ones. Among the 7 bulge dominated galaxies, 
only two show faint emission lines, the remaining having colors and D(4000) 
indices typical of old stellar population. This is consistent with a 
scenario in which most
of the luminous galaxies later than S0 were experiencing enhanced star
formation at z$>$0.5, while most of the luminous E/S0 were still
quiescent. The apparent absence of red quiescent objects beyond z = 0.8 is
likely due to selection effects (see section 2.4.2) and cannot
be simply taken as evidence for evolution of E/S0 galaxies below z = 1.
\begin{figure}[tbp] \label{f5}
\plotone{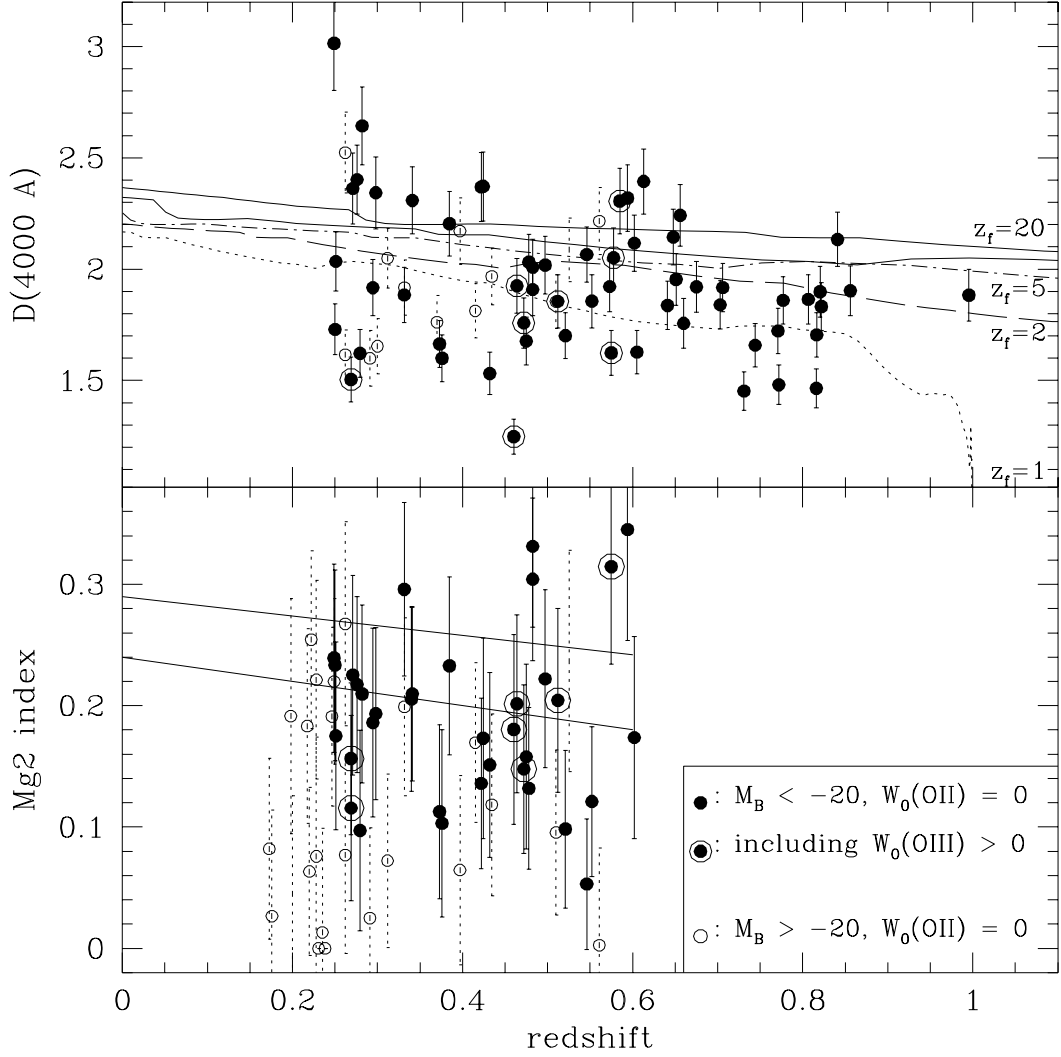}
\caption[]{{\it (top)}: Distribution of the D(4000) index against redshift 
for the 59 $M_{B}< -20$ quiescent galaxies with 0.35$< z <$ 1 (full dots:
$W_{0}(OII)=$ 0; circled dots: $W_{0}(OII)=$ 0 and $W_{0}(OIII)>$ 0). Less 
luminous galaxies are represented by open dots.
Single burst models from BC93 are drawn with various redshift of
formation (see text), assuming $H_{0}=$50.
{\it(bottom)}: Mg$_2$ index against redshift for 59 quiescent galaxies (z$<$ 0.6)
with $M_{B}< -20$ (same symbols as in top).
The lines show the expectations from the Buzzoni et al. (1992) models
for a single burst with $\Omega=$1 and a zero cosmological constant, 
for $H_{0}=$50 (top) and $H_{0}=$100, respectively.}

\end{figure}

Figure 7 shows that the quiescent galaxies define the red envelope of
galaxies for all color indices, with the notable exception of the
Balmer index D(3550-3850). The red envelope of the D(4000) and the
D(3250-3550) indices decrease with  redshift by nearly the same amount,
0.5 magnitude from z = 0.5 to z = 1. However these numbers could be
affected by our failure in detecting the reddest and faintest quiescent
galaxies at high redshift.

Figure 20(upper) shows the distribution of the D(4000) index against 
redshift for 59 quiescent galaxies with $M_{B} < -20$ galaxies (among
256 galaxies between z = 0.23 and z = 1 for which the D(4000) index has
been measured). The lines shown are from BC93 models (single burst),
assuming $H_{0}= $50, and showing various redshifts of formation. The
top two solid lines are for $z_{f}=$ 20, and show the effect of varying
$q_{0}$ (upper: $q_{0}=$ 0.1, lower: $q_{0}=$ 0.5). Assuming
$H_{0}=$100 would move these curves down, reducing the
predicted D(4000) index by a factor 1.1.  At all redshifts, a large
fraction (66\%) of quiescent galaxies lie well below the top lines,
which can be easily understood if they have experienced star formation
events at more recent epochs than z = 2. A few of them have peculiar
emission lines properties and show [OIII] 5007 emission (only detectable 
below
z = 0.7, see section 8.4). 

In view of the good agreement at z$>$0.5 between the fraction of
quiescent galaxies with that of E/S0, the $\sim$33\% of quiescent
galaxies with D(4000) smaller than expected from an instantaneous burst
at $z_{f}$= 1 are likely associated with galaxies with types later than
S0. We notice a possible color evolution of the quiescent galaxies
($W_{0}(OII)$$=$0) between z = 0.5-0.7 and z = 0.7-1. Between z = 0.7
and z = 1, six of the fourteen galaxies (42\%) have restframe
$(U-V)_{AB}$ colors bluer than Sbc while only two out of eighteen with
0.5 $<$ z $<$ 0.7 (11\%) are this blue. This confirms
that some of the quiescent galaxies experienced relatively recent star
formation prior to z = 1. These numbers are likely affected by our
failure in detecting the reddest and faintest quiescent galaxies at
high redshift.

   We have compared the distribution of quiescent objects in the continuum
index diagrams (Figures 15 and 16) in the two redshift bins. At least
six z $<$ 0.7 galaxies are redder than the reddest BC93 stellar tracks
in these diagrams, consistent with $A_{V}\sim$1, but there are no convincing
cases at z$>$ 0.7. This might be an indication of a decrease of the 
average extinction in quiescent galaxies at $z > 0.7$.

It has long been discussed whether the age of high-z ellipticals could
be in conflict with cosmology (Hamilton 1985; Lilly 1988,
and references therein).  There are 12 quiescent galaxies, all
at z $<$ 0.7, which appear older than predicted by any cosmological
models.  Among these objects six appear, from their location in the
Balmer/D(4000) and UV/Balmer diagrams (Figures 15 and 16), to be likely
contaminated by dust. Significant amounts of dust can easily provide
sufficient reddening of the D(4000) index, e.g., an extinction of
$A_{V}$=1 would increase the D(4000) index by a factor 1.08, which
would be enough to account for most of the observed excess.  We have
carefully examined the spectra of the remaining objects and find only
one convincing case of a galaxy 
(CFRS14.1348 at z = 0.613) which shows a
significantly higher D(4000) index than predicted by the usual
cosmologies.  The other five galaxies have either noisy or contaminated
spectra,  or have poor agreement between spectroscopy and photometry.
The spectrum of CFRS14.1348 should be investigated further to see if it
really is discordant.

\subsection{The Mg$_2$ index versus redshift}

The Mg$_2$ index is known to be a very good indicator of metallicity in
elliptical galaxies (e.g., Buzzoni et al. 1992) and is also dependent
on stellar temperature and gravity. The latter author has calibrated
this index as a function of metallicity for a population of coeval
stars using a single stellar population model, assuming a very high
redshift for the initial burst.  Figure 20 (lower panel) shows the
Mg$_2$ index against redshift for the CFRS galaxies which are
apparently early type (objects as red as local E/S0, defined from the
$V-I$ color vs redshift diagram), and an absence of emission lines
([OII] 3727, [OIII] 4959, 5007). The two solid lines in the diagram
show the Buzzoni et al. (1992) models (for a single burst) for
$\Omega=$1 and a zero cosmological constant. The upper line corresponds
to $H_{0}=$50 while the bottom one corresponds to $H_{0}=$100. It is
interesting to notice that most of the points are found below the
$H_{0}=$100 line. Other cosmologies ($\Omega$=$\Lambda$=$0$;
$\Omega$=$0.1$ and $\Lambda$=$0.9$) predict higher Mg$_2$ index and are
even less attractive. This result should be taken with caution since
the error bars on Mg$_2$ are large. However, it provides some support
for the suggestion that a large fraction of the quiescent galaxies have
experienced bursts of star formation at more recent epochs than z = 2.

\section{PECULIAR OBJECTS}

\subsection{Blue nucleated galaxies}

We investigate here the spectral properties of the sample of 32 CFRS
galaxies which have been observed with HST in Cycle 4 (CFRS IX). In
that paper, we found that 30\% of the galaxies, referred to as ``blue
nucleated galaxies",  are dominated by blue compact components.  These
components occur most often in peculiar/asymmetric galaxies.  Figure 21
(upper left) shows the average spectrum of the 10 blue nucleated
galaxies. The latter have emission line ratios spanning the whole range
of CFRS emission-line galaxies (see Figure 12b), and none have
Seyfert2-like spectra. They apparently show higher star formation rates
than the rest of the CFRS IX sample, since their average [OII] 3727 
luminosity
($\sim$ $3\times10^{41}$ erg s$^{-1}$) is twice that of the other
emission-line galaxies. They also include the galaxy with the highest
[OII] 3727 luminosity (CFRS14.0972, $L_{OII}= 1.2\times10^{42}$ 
erg s$^{-1}$)
in the subsample. Their UV and UV-B color indices (D(3250-3550) and
D(3550-4150)) are also slightly bluer (by $\sim$ 0.2 mag) on average
than the other galaxies.
\begin{figure}[tbp] \label{f5}
\plotone{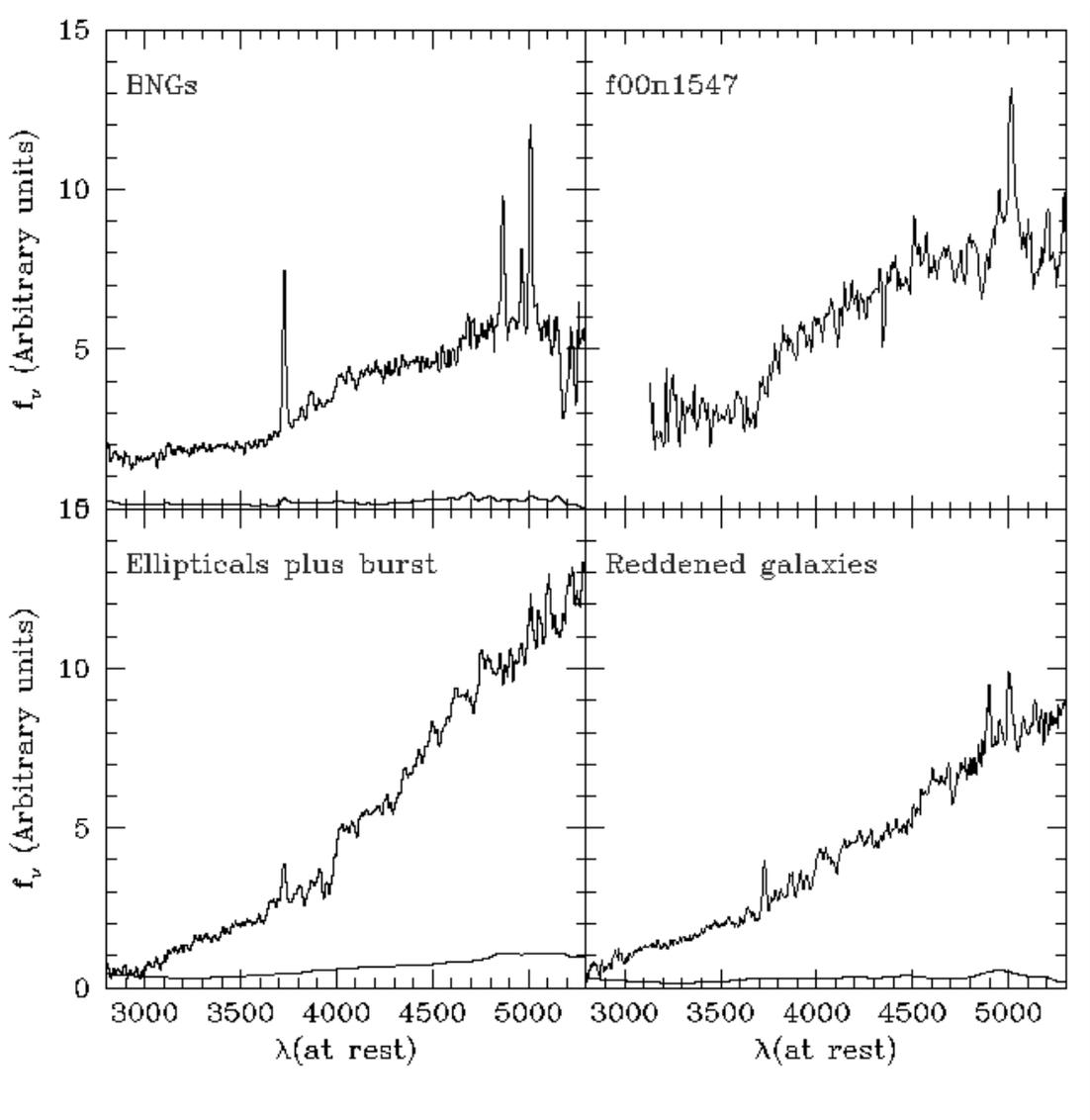}
\caption[]{ {\it (a, top-left)}: Average spectrum of the 10 blue nucleated
galaxies (BNG) from the CFRS IX subsample. It shows stronger [OII] 3727
emission than the rest of the sample (($W_{0}(OII)=$ 40\AA). {\it (b, top
right)}:  Spectrum of CFRS00.1547 which shows [OIII] 4959 and 5007
emission lines, a large Balmer break but no 4000\AA\ break). {\it (c, bottom
left)}: Average spectrum of the 13 galaxies classified as
``elliptical+miniburst" from their continuum and emission line
properties (see text). It shows a nice 4000\AA\ break and a moderate
[OII] 3727 emission line ($W_{0}(OII)=$ 10\AA). {\it (d, bottom right)}
Average spectrum of the 10 galaxies classified as ``young
heavily-absorbed" from their continuum and emission line properties
(see text). These galaxies have UV properties redder, or as red, as old
stellar populations, while their visible properties are typical of
stellar populations younger than 1 Gyr. The average spectrum is almost
a straight line from 3000 to 4500\AA\ and shows moderately strong [OII]
3727 emission ($W_{0}(OII)=$ 20\AA). Residuals 
(1$\sigma$ deviation) are shown in the bottom of each panel.}

\end{figure}

\subsection{Post-starburst galaxies}

We have adopted a similar definition for this category of objects as
Zabludoff et al. (1996), i.e., objects without emission lines but with
prominent Balmer absorption lines (average equivalent width of
$H\delta,\gamma,\beta$ larger than 5.5\AA). Since our spectral
resolution does not allow accurate estimation of the absorption line
equivalent widths, we used the BC93 models to calibrate the Balmer
index with Balmer line strength. We find that objects with
log(D(3550-3850)) $> $0.2 would likely have equivalent widths of
5-7\AA, for most star formation scenarios.  Inspection of Figure 15
shows ten such objects among 204 galaxies (with M$_{B} < -20$, defined
[OII] 3727 and continuum indices), corresponding to a fraction of $\sim$ 5\%,
higher than that  found locally ($\sim$ 0.2\%) by Zabludoff et al.
1996.  The difference could be a result of the very severe limitation
on $W_{0}(OII)$ ($<$ 2.5\AA) set by Zabludoff et al., since our
detection limit is more likely around 5 to 10\AA. There is also a
suggestion of a decrease of the fraction of post-starburst galaxies with
redshift, since only one lies at z $>$ 0.7.

\subsection{Emission-line galaxies with red UV colors}

We were intrigued by the presence of galaxies which have emission lines 
but with D(3250-3550) indices indicative of very old stellar populations
(see Figure 16). Among 210 galaxies having z $>$ 0.45 (and with M$_{B}
< -20$, [OII] 3727 and relevant continuum indices defined), there are 23
emission-line galaxies with D(3250-3550) $>$1.4.  This value
corresponds to a relatively old stellar population (age 5 Gyr with BC93
code, or 3 Gyr with BC95 code).  This population can be subdivided by
location in the Balmer-D(4000) diagram. One population of galaxies (13
objects) is located on the old stellar tracks (ages $\sim$ 3-5 Gyr),
the 10 remaining galaxies being distributed along the young stellar
track (ages $<$ 1 Gyr). We believe that the first population could be
identified as ellipticals which are experiencing extremely modest
bursts (involving $<$1\% of the mass according to our simulations),
while the second one is likely a population of young, heavily-absorbed
galaxies ($A_{V}>$ 2).  Figure 21 shows the combined spectrum of these
``elliptical + mini burst"  and ``vigorously star-forming,
heavily-absorbed" galaxies. The latter objects have a UV spectrum from
3000 to 3900\AA\ similar or redder than that of a 5 Gyr elliptical,
while the red part of their spectra (above 4000\AA) is much bluer than
an elliptical.  They also often show H$\beta$ and [OIII] 5007 in their
spectra. Young and heavily absorbed galaxies represent 8\% of the
$M_{B}< -20$ galaxy population at z $<$ 0.7, while only one of them
lies at z $> $ 0.8.  Their apparent [OII] 3727 luminosities are relatively
low ($L_{[OII]} < 3\times10^{41}$ erg s$^{-1}$) consistent with the
fact that high extinction would affect the [OII] 3727 luminosity.

\subsection{Galaxies with peculiar emission lines.}

We have looked for objects having [OIII] 5007 lines but no [OII] 3727
lines.  We identify six such objects after a careful analysis of their
spectra.  Figure 21 (upper right) shows the spectrum of CFRS00.1547
which is a good example. The six galaxies have redshifts ranging from z
= 0.2 to z = 0.6 and are relatively luminous ($M_{B}$ $\sim -20.5$).
For the two lowest redshift objects, we have detected a prominent
$H\alpha$ line, while all of them have H$\beta$ in absorption. HII
regions in these objects might be heavily affected by extinction, while
the continuum and the H$\beta$ absorption line indicate that these
galaxies have experienced strong starbursts in the recent past (few
tenths of Gyr). An interesting possibility is thus that most of the gas
might have been consumed in this burst and the volume of gas available
for ionization is much smaller than the volume that the OB associations
are capable of ionizing (see e.g., McCall et al. 1985). Another
alternative might be that these objects are related to AGNs.

\section{DISCUSSION}

The most obvious interpretation of the evolution of the properties of
CFRS galaxies from z = 0.2 to z = 1 (section 3.3) is in terms of a
global increase with redshift of the star formation rate in disk
galaxies. Such a scenario would account for the evolution of the [OII] 3727
emission line properties as well as for the ``blueing" of the most
luminous emission-line galaxies beyond z = 0.5.

An interesting question raised by Cowie et al. (1995) is to estimate
what fraction of present-day stars have been formed since z = 1, based
on the observed [OII] 3727 luminosity densities.  Star formation rates
have been calibrated by K92 from a sample of local galaxies covering
the full range of morphological type observed (from Sa to Irr).
Assuming that the K92 calibration of the SFR is valid for all $M_{B} <
-18.5$ CFRS galaxies at all redshifts, we then compute, for an age of
the Universe of 13 Gyr ($H_{0}$=50 and $q_{0}$=0.5), that the stellar
mass formed between $0 \le z \le 1$ would be 3.87 $\times10^{8}
M_{\odot}$  and this would exceed the present day stellar mass (Figure
6), assumed to be $3\times10^{8}h_{50} M_{\odot}$ per $Mpc^{-3}$ (Cowie
et al. 1996; Glazebrook et al. 1995). This is clearly unacceptable:
there are many ``old'' galaxies at $z \sim 1$ (see section 7) and there
are many star-forming galaxies beyond z = 1 (this paper, Cowie et al.
1995, Steidel et al.  1996). One of the assumptions must be wrong. As
discussed in CFRS XIII, the cosmological uncertainties involving volume
elements and cosmological timescales more or less cancel out for an
assumed total age of the Universe, so changing the cosmological
parameters is an unattractive solution (a value of $H_{0}=62$ and
$q_{0}=0.1$ would slightly reduce the production of present day stars
to 3.3 $\times10^{8} M_{\odot}$). There is a good agreement between
Cowie et al. (1996) and Glazebrook et al. (1995) in their estimation of
the (local) cumulative luminosity density in K (they find
$2.5\times10^{8}h_{50}L_{\odot}$ per $Mpc^{-3}$ and
$3\times10^{8}h_{50} L_{\odot}$ per $Mpc^{-3}$, respectively). And the
current models of galaxy spectra predict a mass to infrared ratio very
close to the unity and almost insensitive to the galaxy color and type
(Charlot, 1996b).

Two of the most likely possibilities that might affect the [OII] 3727
luminosity--SFR relationship at high redshifts are differing IMFs or
evolving metallicities.  If the IMF were richer in massive stars in higher 
redshift
galaxies then, for the same amount of UV luminosity density (or [OII] 3727
luminosity density), they would produce fewer long-lived stars. In this
case the most luminous galaxies at 1$\mu$ would not necessarily be
associated with the most massive galaxies since the near-IR light could
be still dominated by short-lived stars. The compact narrow
emission-line galaxies described by  Guzman et al. (1996) might be
examples, since they are luminous but evidently not massive.  It should
be remembered that the IMF required to fit the spectrophotometric
properties of local disks is estimated to be rather rich in massive stars, 
likely following a Salpeter law rather than a Scalo law (Kennicutt et al.
1994). To explain the observed properties of CFRS galaxies, a rather
extreme IMF at high redshifts would be required.

Alternatively an evolution in the average metallicities of the stars
and in the degree of dust extinction could change the [OII] 3727--SFR
relation. We have seen how our data suggest that the metallicities of
some of the most luminous galaxies in [OII] 3727 at high redshift may
be substantially lower than observed in luminous galaxies today, and
the attendant effects, particularly on dust absorption, may contribute
to many of the observed phenomena.  We have analysed the effects of
metallicity on the continuum indices by comparing the BC95 model to one
with extremely low metallicity ($Z=0.02Z_{\odot}$, Bruzual \& Charlot
1996). Rest-frame $(U-V)_{AB}$ colors are much smaller at lower
metallicities (from 0.5 to 1 mag for ages larger than $5\times10^{8}$
years). For some spectral indices but not all, lower metallicities
correspond to bluer indices at all ages. This effect can easily reach
several tenths of magnitude for the D(4000) and D(41-50) indices. For
only one index, the Balmer index D(3550-3850), the decreasing
metallicity slightly reddens the continuum. On the other hand,
metallicity has almost no effect for the near UV index, D(3250-3550),
for any age. A decrease of the metallicity with the redshift appears to 
be grossly consistent with the reported redshift evolution of the colors. 
An increase of the fraction of galaxies with lower
metallicities than solar values would also account for the increasing
dispersion of the relation between D(4000) index and $W_{0}(OII)$ or
[OII] 3727 luminosity (Figure 9).

We also detect a significant population of galaxies at high redshift
(1/3 of the sample z$>$0.7) with indications of low metal abundances 
($Z/Z_{\odot} <$
0.2), which could be almost transparent galaxies, with absorption even
lower than the average value of present-day irregular galaxies (section
6.3). At z $>$ 0.7, the restframe $(U-V)_{AB}$ colors of the most
luminous emission-line galaxies ($M_{1\mu} < -22$) shift towards those
of irregular galaxies. It appears that the metal-deficient population
of galaxies at high redshift may represent the extreme cases of a
general decrease of the metallicity and absorption in emission-line
galaxies from z = 0.4 to z = 1.

Lower average intrinsic absorption and metallicities in the past seem
to provide a natural explanation of the apparent over-production of
long-lived stars. A decrease of the absorption from the value
($A_{V}$=1.26) adopted by K92 for ``normal" local galaxies, closer to
the Gallagher et al. (1989) value ($A_{V}$=0.57) for local irregulars,
would increase the apparent luminosity densities at 2800\AA\ and in the
[OII] 3727 emission line at high redshift by a factor 3.8 and $>$ 2.8
respectively, compared to observed increases of factors of about 15 and
11 respectively. Locally it has been widely assumed (see Gallego et
al.  1995) that galaxies have average properties similar to those in
the K92 sample, which span the whole range of morphological types from
Irr to Sa. From a sample of irregular galaxies, Gallagher et al. (1989)
find a calibration of the star formation rate 5 times lower than K92
(for the same [OII] 3727 luminosity). Approximately 60\% of the difference is
due to the lower average extinction found in the Gallagher et al.
sample, and K92 quote that 25\% is related to the different IMF used in
the two studies.  The effects of IMF and extinction may in fact be
related:  in a recent study of star clusters von Hippel et al. (1996)
find that, with decreasing metallicities, the stellar luminosity
function extends further towards the brighter end.

In summary, there seems to be good evidence that the average metallicity
of galaxies was considerably lower at higher redshifts, that the
intrinsic absorption was also much lower, and that the combination of
these can contribute to much of the evolution observed in the CFRS spectra.
An important test of this idea will be to measure the mid- and far-infrared
luminosities of these high redshift galaxies.

As a simple demonstrative model, we assume that the physical conditions
(opacity, metallicity and IMF) which govern the SFR calibration are
linked to a single parameter, the restframe $(U-V)_{AB}$ color.
Lower values for the latter are likely associated with higher values of
the birth rate parameter and to higher opacities and metallicities, as
in present-day irregular galaxies. We further assume the Gallagher et
al. [OII] 3727--SFR calibration for galaxies having colors of irregulars
($(U-V)_{AB}= 0.5$) and the K92 SFR calibration for galaxies with
colors of Sbc ($(U-V)_{AB} = 1.3$), the latter having average colors of
the K92 sample. For other galaxies, these two points allow us to
interpolate the SFR calibration, following the formula:\\

SFR =$L_{[OII]}$ * 5 $\times10^{-41}$ * $10^{0.559*((U-V)_{AB} - 1.3)}$ \\
\begin{figure}[tbp] \label{f5}
\plotone{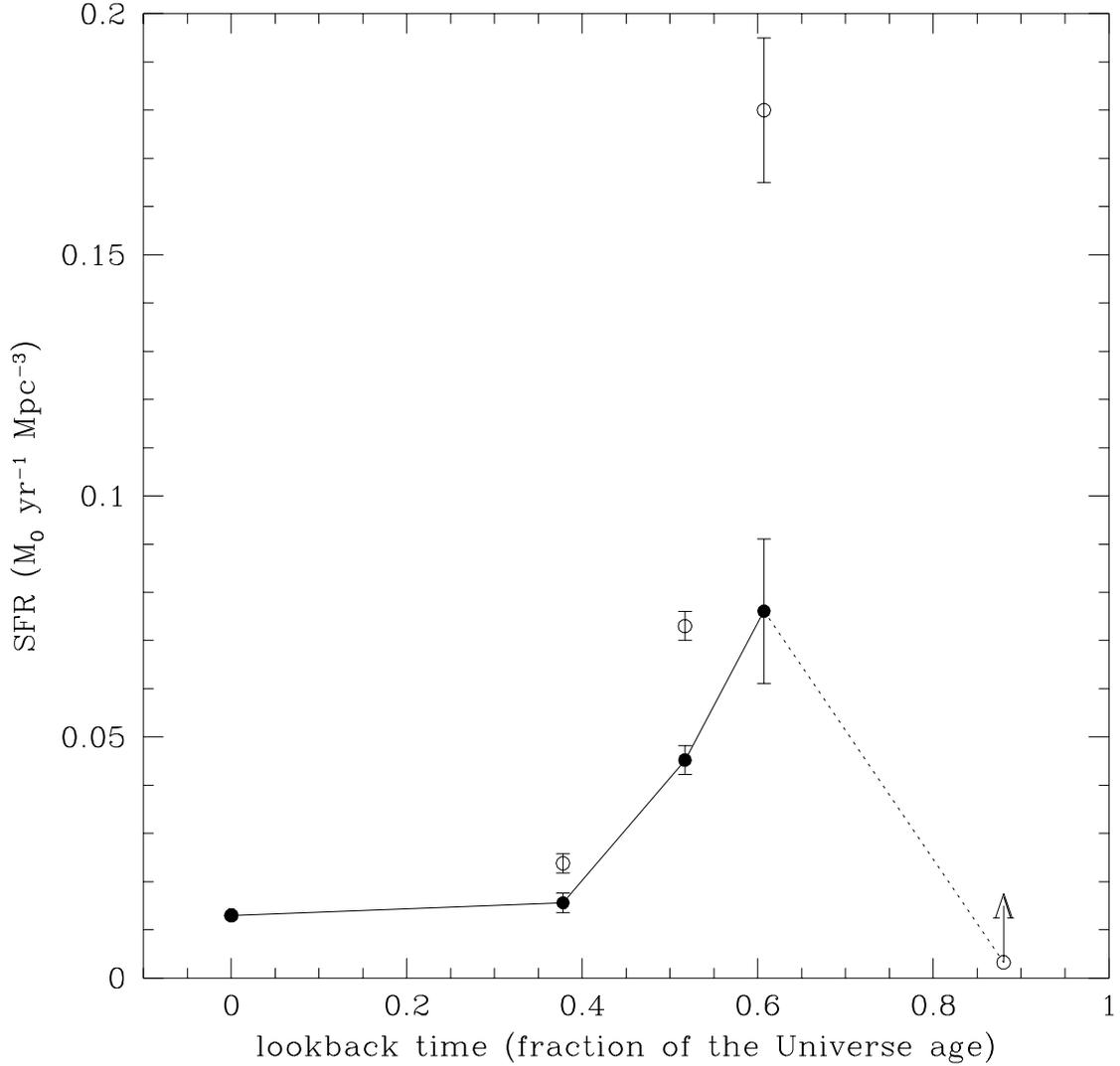}
\caption[]{Star formation history predicted by our simple model in
which we assume that the physical properties (opacity, metallicity and
IMF) of galaxies at all redshift depend on the restframe $(U -
V)_{AB}$ color as they do in the local Universe. Full dots represent
the prediction from the models, the dots at lookback-times 0\% and
87\% (z = 3.2) of the present age of the Universe are from Gallego et
al. (1995) and Steidel et al. (1996). Open circles represent the same
values, if  no evolution of the physical properties has occurred from z = 0
to z = 1. The latter values would produce more long-lived stars since z = 1,
than observed locally.}

\end{figure}

Figure 22 shows the predicted evolution of the SFR density vs lookback
time (scaled in units of the present age of the Universe).  The SFR
density shows little evolution from z = 0 to z = 0.375 (lookback time 
38\% of the age of the Universe), and then strongly increases towards
larger lookback times. In this model, the fraction of present-day stars
formed since z = 1 (75\%) happens to be slightly higher than the
elapsed fraction (66\%) of the age of the Universe.  This suggests that
the overall SFR rate should show a turnover not far beyond our present
limit at z $\sim$1. In this context it is worth noting the relatively
low levels in the overall SFR found at z$\sim$3.2 by Steidel et al.
(1996). The average absorption ($A_{V}$) would have decreased by
$\sim$0.6 mag (50\%) from z = 0 to z = 0.85. At low redshifts
(z $<$ 0.5), the [OII] 3727 luminosity density is dominated by less 
massive galaxies, while
its increase beyond z = 0.5 is essentially due to the increasing
contribution of luminous galaxies due to both increased star formation
and decreased dust opacities. At z $> $0.7, a significant fraction of
galaxies would have only recently formed their first generation of
stars, and would have significantly lower metallicity and absorption
than present-day galaxies.
  
Continuum luminosities might be also affected by a reduction of the
absorption at higher redshift. Locally, de Vaucouleurs (1995) quotes
$A_{V}$$\sim$1 for the average intrinsic absorption of a local Sbc
galaxy.  A change in average absorption of $A_{V}$ by 0.6 magnitudes
would, on its own, double the luminosity density at B, and thus account
for about a half of the observed increase in overall luminosity density
that was reported in CFRS XIII, reducing the exponents $\alpha$ by
about 1. It should also play a role in the evolution of the B
luminosity function as well as in the observed disk brightening,
especially beyond z = 0.7 (expected effect as high as 0.9 mag).

\section{CONCLUSIONS}

The main spectrophotometric properties of the CFRS sample of 
field galaxies and the evolution of these properties can be briefly 
summarized as follows:\\
-- there is strong change in the comoving luminosity density of [OII] 3727
by a factor 8.3$\pm$4.3 from z = 0.375 to z = 0.85, but relatively modest
changes in the restframe equivalent widths $W_{0}[OII]$ and the
D(4000) breaks exhibited by individual galaxies, and essentially no
evolution in purely ultraviolet colors.\\
-- a large fraction of emission-line galaxies (40\%) contain a
substantial population of A stars indicating that the star-formation 
is proceeding in a sustained fashion.\\
-- there is a significant fraction of emission-line galaxies at high
redshifts (25 to 30\% at z$>$ 0.7) with very small D(4000) indices,
suggesting metallicities much lower than solar values ( $Z/Z_{\odot} <$
0.2), in which the star formation may have started several $10^{8}$
years previously in an almost primordial medium.\\
--  the ionization parameters of HII regions in these galaxies were
generally higher at higher redshifts.\\

There are several reasons to believe that emission-line galaxies are
more transparent at higher redshift.  At z$>$ 0.7, the colors of
the emission-line galaxies which are luminous at 1$\mu$ move closer to
those of local irregulars, and the contribution of these galaxies to
the [OII] 3727 luminosity density becomes important. We also observe a
significant increase of the ionization parameter of HII regions and in
addition, a population of galaxies with small 4000 \AA\ breaks, the
D(4000)-deficient objects, appears at z $>$ 0.7.  These likely have low
metallicities.

It appears that the local Kennicutt (1992) relationship for the calibration
of the SFR in terms of the [OII] 3727 luminosity cannot be used directly at
high redshifts.  Not least, it leads to a serious over-production of
stars relative to those seen locally.  We suggest that effects
concerned with the evolution of metallicity and intrinsic absorption of
galaxies must be taken into account. The evolution of the internal
obscuration between the hot stars and gas in HII regions may also be a
factor in the evolution of the strength of [OII] 3727 and might be the
source of the evolution of emission-line properties in HII galaxies
(section 4.3).  Such effects may also contribute to the observed
brightening of disk galaxies (CFRS IX). 
 Changes in the opacity/metal 
abundance in galaxies may be also related to changes of the IMF,
and a steepening of the latter in the past may also contribute to
solve the star over-production problem. 

A simple, single parameter, model can be used to demonstrate how the
calibration of the [OII] 3727---SFR relation might change with epoch,
and what effect this would have on conclusions drawn from the galaxies
in the CFRS sample.  In this model the calibration between [OII] 3727
luminosity and star-formation rate is assumed to depend only on the
observed restframe (U-V) color.  This model broadly accounts for most
of the observed changes with the redshift of the spectrophotometric 
properties seen in individual CFRS galaxies, and can account for about 
a half of the
increase that is observed in the 2800\AA\ and [OII] 3727 comoving 
luminosity
densities, in the luminosity function of blue galaxies and in the
 characteristic surface brightnesses of disks at high redshift.

The indication of low metal abundances in a significant fraction of 
z$>$0.7 field galaxies could be consistent with the very low abundances 
found in the
damped $Ly\alpha$ systems from z = 0.7 to z = 3.4 (Smith et al.  1996),
as well as with the weakness of the absorption features in the spectra
of z$> $3 galaxies identified by Steidel et al. (1996). 

In the future, detailed fits of the galaxy spectra from UV to 1 micron
will be made to examine the effects of reddening.  Measurements of the
$H\alpha$ line redshifted into the near IR and scheduled ISO observations
of the CFRS fields will provide invaluable results about the evolution
of dust properties in field galaxies up to z = 1.  In addition,
spectroscopic observations of CFRS galaxies beyond 8500\AA\ will bring
several emission lines into the spectroscopic window that will enable
the evolution of the properties of HII regions to be followed to higher
redshifts. 
 Analysis of HST images of a large sample of CFRS galaxies will
also provide new tests on the conditions (merging, distribution of HII 
regions, dust etc...) which drive the star formation in $z<$1 galaxies.

\acknowledgements 
This paper has benefitted from many interactions with many colleagues.
Special thanks to S. Charlot (and to G.  Bruzual) who have shared with
us unpublished results of their most up-to-date GISSEL code. Thanks
also to D. Alloin, P. Jablonka and E. Bica who have allowed us to use
their library of star clusters. D. Pelat has introduced us to the use
of the MEASURE code. M. Serote has provided  a library of star spectra.
Many interactive discussions with most of the above-mentioned people
should be acknowledged as well as with G. Stasinska, C. Leitherer, and
P. Vettolani, who has also kindly calculated for us the fraction
of bright emission line galaxies with significant star formation in the
ESP sample. This paper has benefitted a lot from the comments and 
suggestions of an anonymous referee. We thank the directors of the 
CFHT and the two allocation
time committees (CFGT and CTAC) for their support of the CFRS project.
SJL, GM-O and DJS researches are supported by the NSERC of Canada and we 
acknowledge
travel support from NATO and from the project ECOS/CONICYT C92U02.

\end{document}